\documentclass[%
 reprint,
 superscriptaddress,
 amsmath,amssymb,
 aps,
]{revtex4-2}
\usepackage{tikz}
\usetikzlibrary{quantikz}
\usepackage{graphicx}% Include figure files

\usepackage{dcolumn}% Align table columns on decimal point
\usepackage{bm}% bold math
\usepackage[colorlinks=true,linkcolor=blue,urlcolor=black,citecolor=blue,bookmarksopen=true]{hyperref}
\usepackage{amsmath,physics}
\usepackage{xcolor}
\usepackage[caption=false]{subfig}

\usepackage{cancel}
\begin{document}

\preprint{APS/123-QED}

\title{Quantum simulation of molecular response properties}% Force line breaks with \\

\author{Ashutosh Kumar}
\email{akumar1@lanl.gov}
\affiliation{Theoretical Division, Los Alamos National Laboratory, New Mexico, 87545, USA}
 \author{Ayush Asthana}
 \affiliation{Department of Chemistry, Virginia Tech, Blacksburg, VA 24061, USA}
 \author{Vibin Abraham}
 \affiliation{Department of Chemistry, University of Michigan, USA}
 \author{T. Daniel Crawford}
 \affiliation{Department of Chemistry, Virginia Tech, Blacksburg, VA 24061, USA}
%\affiliation{%
 \author{Nicholas J. Mayhall}
\affiliation{Department of Chemistry, Virginia Tech, Blacksburg, VA 24061, USA}
\author{Yu Zhang}
\email{zhy@lanl.gov}
\affiliation{Theoretical Division, Los Alamos National Laboratory, Los Alamos, NM 87545, USA}

\author{Lukasz Cincio}
\affiliation{Theoretical Division, Los Alamos National Laboratory, Los Alamos, NM 87545, USA}
\author{Sergei Tretiak}
\affiliation{Theoretical Division, Los Alamos National Laboratory, Los Alamos, NM 87545, USA}
\affiliation{Center for Integrated Nanotechnologies, Los Alamos National Laboratory, Los Alamos, NM 87545, USA}
\author{Pavel A. Dub}
\email{pdub@lanl.gov}
\affiliation{Chemistry Division, Los Alamos National Laboratory, Los Alamos, NM 87545, USA}

 %This line break forced with %\textbackslash\textbackslash
%}%


\begin{abstract}
Accurate modeling of the response of molecular systems to an external electromagnetic field is challenging on classical computers, especially in the regime of strong electronic correlation. 
In this paper, we develop a quantum linear response (\texttt{qLR}) theory to calculate molecular 
response properties on near-term quantum computers.
Inspired by the recently developed variants of the quantum counterpart of equation of motion (\texttt{qEOM}) theory, the \texttt{qLR} formalism employs ``killer condition'' satisfying excitation operator manifolds that offers a number of theoretical advantages along with reduced quantum resource requirements. 
We also used the \texttt{qEOM} framework in this work to calculate state-specific response properties.
Further, through noise-less quantum simulations, we show that response properties calculated using the \texttt{qLR} approach are more accurate than the ones obtained from the classical coupled-cluster based linear response models due to the improved quality of the ground-state wavefunction obtained using the \texttt{ADAPT-VQE} algorithm.
% recipe for evaluation of response properties (qLR) on a quantum computer based on response theory. We further used qLR theory along with earlier developed variants of qEOM theory to test these methods by computing response properties with a ground state molecular wavefunction prepared using a VQE algorithm on a statevector simulator.
% The results obtained are of better quality than the ones obtained using the classical linear-response coupled-cluster methods, which are considered to be highly accurate,  taking advantage of an improved quality of ground state wavefunction obtained using VQEs.
\end{abstract}

\maketitle

\section{Introduction}
The field of quantum chemistry has made significant progress in recent decades in the accurate numerical simulation of electronic properties of a wide range of molecules and materials~\cite{review_quantum_chem,evangelista2018perspective,liu2021relativistic,motta2017towards,chan2011density,mardirossian2017thirty,liu2020mapping}. 
However, a number of challenges still remain. 
The computational complexity of accurate electronic structure methods continues to be quite high, especially when strong electron correlation effects are involved, where the numerical evaluation of the ground and low-lying excited states of the molecular Hamiltonian may scale factorially with respect to the system size~\cite{FeMocopaper}. 
%In the recent decade, a number of computationally practical reduced-scaling approaches have been developed~\cite{VRG:riplinger:2016:JCP,VRG:kumar:2020:JCP} but these techniques mostly apply to chemical problems dominated by weak electron correlation effects. 
With the advent of quantum devices that exploit the quantum properties of superposition and entanglement, one can map the exponentially increasing Hilbert space to a linearly scaling number of qubits~\cite{nielsen2002quantum}. 
The quantum hardware in the NISQ era, however, suffers from a number of challenges like limited qubit connectivity, significant gate-error rates, short coherence times, etc., which prevents us from realizing the promised ``quantum advantage". 
The variational eigensolver (VQE) method~\cite{peruzzo2014variational} attempts to overcome some of these limitations by ensuring shallow quantum circuits through a variational optimization of the quantum circuit parameters. 
This has allowed for the development of a number of quantum algorithms for the simulation of molecular ground~\cite{grimsley2019adaptive,tang2021qubit,peruzzo2014variational,kandala2017hardware,huggins2020non,lee2018generalized,ryabinkin2018qubit,o2016scalable,nam2020ground,mccaskey2019quantum,hempel2018quantum,gao2021applications,smart2021quantum,Meitei2020,tkachenko2021correlation,clusterVQE,asthana2022minimizing} and excited states~\cite{nakanishi2019subspace, xie2022orthogonal, chan2021molecular, higgott2019variational, colless2018computation,mcclean2020decoding,Takeshita2020,mcclean2017hybrid, qEOM2020,asthana2022equation}. 
Aside from the VQE method, algorithms based on quantum phase estimation~\cite{abrams1999quantum, aspuru2005simulated}, adiabatic state preparation~\cite{farhi2001quantum, babbush2014adiabatic}, and Krylov subspace generation~\cite{qdavidson,GarnetNP2020,jctc.9b01125} have also been developed for molecular simulations. These techniques are more suitable for the era of fault-tolerant quantum computing.

Most of the quantum computing applications in chemistry have been focused on the estimation of ground and excited-state energies with limited attention to molecular response properties.
As the name suggests, these properties capture the response of the electric dipole moment of a molecule to an external field.
For example, the molecular polarizability is defined as the first-order response of its electronic charge distribution to an external electric field. 
Polarizabilities are at the origin of many chemical phenomena including electron scattering~\cite{lane1980theory}, electronegativity~\cite{vela1990relationship}, softness and hardness~\cite{Datta94hardness}, and they play an important role in biological processes such as protein-ligand binding ~\cite{guo2000many}.
When strong electric fields are involved, as in the case of lasers, higher-order response properties such as hyper-polarizabilities (second-order) also become significant. These quantities, for example, define the suitabilitity of materials for nonlinear optical applications~\cite{wagniere1993linear}.
Chiroptical properties are another class of response properties that finds several applications in the pharmaceutical industry.
More than half of the drugs currently in use are chiral~\cite{NguyenChiral}, i.e. the molecular structure of these drugs has a unique three-dimensional handedness and thus exists in the form of left- and right-hand stereoisomers, also known as enantiomers. 
Within a chiral environment, the chemical properties of the enantiomers can be drastically different. This underscores the importance of understanding the structure-activity 
relationship of these compounds \cite{crawford2006ab}.
Optical rotation, which refers to the rotation of the plane of plane-polarized light as it passes through a chiral medium, is a useful tool in determining the absolute configuration of chiral molecular systems. 
Just like polarizabilities, optical rotation can be characterized as the first-order response of the electric dipole moment, but with respect to an external magnetic field.

The exact treatment of these response properties can be carried out by the sum-over-states (SoS) formalism~\cite{koch1990coupled}, which involves explicit evaluation of all the excited states associated with the molecular Hamiltonian. 
Consequently, implementing the SoS approach for even medium-sized molecular systems can be challenging. An alternative approach is built on the idea of expanding the perturbed wave functions in a determinantal basis rather~\cite{HirataPolar2005,koch1990coupled,christiansen1998response,pedersen1997coupled} thus, avoiding the explicit determination of the excited states.
In classical quantum chemistry,  coupled cluster (CC) response theory (RT), developed extensively by Koch, J\"{o}rgensen, and co-workers~\cite{koch1990coupled,pedersen1997coupled, christiansen1998response} is one of the most promising approaches in this regard. 
Another popular approach is through the use of equation of motion coupled cluster (EOM-CC) theory introduced by Stanton and Bartlett~\cite{Stanton1993}. Unlike the CC-RT formalism, this method attempts to calculate response properties within the SoS framework based on excited states computed using EOM-CC. Green's function based approaches~\cite{linderberg2004propagators, dreuw2015algebraic} are yet another class of methods which are frequently used to calculate molecular response properties.
It should be noted that excitation energies and transition moments generally also come under the purview of response properties. 
Since both these properties are state-specific, they can be calculated efficiently by both CC-RT and EOM-CC based approaches.

Important recent developments have been made in computing response properties on a quantum computer\cite{huang2022variational,CaiResponse2020,singh2022quantum,teplukhin2021computing,rizzo2022one}. The variational quantum response (VQR) algorithm developed by Huang and co-workers~\cite{huang2022variational} is notable in this regard. The VQR approach transforms the response formalism into an optimization problem that minimizes a cost function using a parameterized quantum circuit to calculate dipole polarizabilities and absorption spectra. 
A number of recently developed quantum excited state methods like subspace-search VQE (SS-VQE)~\cite{nakanishi2019subspace}, the orthogonal state reduction variational eigensolver (OSRVE)~\cite{xie2022orthogonal}, and variational quantum deflation (VQD)~\cite{chan2021molecular, higgott2019variational} operate on similar principles with appropriately designed cost functions for excited state energies. 
Although these methods are promising, they suffer from challenges like increased circuit complexity, and there may be additional challenges finding the global minimum in different cost-function optimization landscapes~\cite{cerezo2021cost}.
Alternatively, excited states can also be obtained by diagonalizing the Hamiltonian in a subspace, just like the classical EOM-CC based approaches. 
Quantum equation of motion (\texttt{qEOM})~\cite{qEOM2020} and quantum subspace expansion (\texttt{QSE})~\cite{colless2018computation,mcclean2020decoding,Takeshita2020,mcclean2017hybrid} methods are popular examples in this regard. 
These methods have the same circuit complexity as the ground state but feature an increase in the number of measurements and higher body reduced density matrix (RDM) requirements. 
%Considering the rapid advances in the measurement technology~\cite{crawford2021efficient, huggins2021efficient,izmaylov2019unitary} this approach seems promising for near-term applications. 
However, the 
\texttt{qEOM} approach does not necessarily satisfy the important ``killer" or vacuum annihilation condition~\cite{prasad1985some,szekeres2001killer} while 
the \texttt{QSE} approach does not guarantee the correct scaling (size-intensivity~\cite{nooijen2005reflections,shavitt2009many}) of energy differences. 
The \texttt{q-sc-EOM} approach developed by us recently~\cite{asthana2022equation} satisfies the ``killer" condition by making use of the self-consistent excitation manifold~\cite{prasad1985some} of Mukherjee. 
Further, it transforms the generalized non-Hermitian eigenvalue problem of \texttt{qEOM} into a Hermitian eigenvalue problem, provides size-intensive energy differences, and is expected to be more noise-resilient compared to other diagonalization-based excited state approaches.

In this work, we develop a quantum counterpart of linear response formalism, namely \texttt{qLR} theory, to calculate molecular response properties such as polarizabilities, optical rotation, etc., on near-term quantum computers. 
We are mainly interested in calculating off-resonance response properties in this work. It should be noted that the damped version of response theory\cite{kristensen2009quasienergy} is used to calculate resonant response properties in classical quantum chemistry and one can easily extend the \texttt{qLR} theory along similar lines to simulate such properties on a quantum computer.
We also make use of the quantum equation-of-motion framework developed in Refs.~\cite{asthana2022equation} and~\cite{fan2021equation} for quantum simulation of state-specific response properties like transition moments and excitation energies. 

This manuscript is structured as follows: Section~\ref{RT} discusses the theoretical formalism for the \texttt{qLR} theory. Section~\ref{VAC} introduces the ``killer condition" and the two types of excitation operator manifolds, namely self-consistent (\texttt{sc}) and projected (\texttt{proj}) operator manifolds that satisfy this condition, and derives the final working equations.
The proposed implementation steps are shown in section~\ref{implementation} while the computational details for all the calculations in this paper are reported in section~\ref{compdet}. Section~\ref{results} discusses the results obtained for H$_2$, LiH, H$_2$O, chiral (H$_2$)$_2$ and linear H$_6$ molecular systems. The key findings of this manuscript are summarized in section~\ref{conclusions}. For completeness, the Appendix (sections \ref{Exp_value} and \ref{EOM}) presents some aspects of linear response theory and the theoretical framework of the \texttt{qEOM} method.
\section{Theory}\label{theory}
\subsection{Linear response theory }\label{RT}
 Molecular response theory captures the interaction of a molecule with an external electromagnetic field based on the time-dependent perturbation theory framework, starting from the time-dependent Schr\"{o}dinger equation
\begin{equation}
    \hat{H}\ket{\Psi_{0}(t)}=\mathrm{i} \frac{d}{d t}\ket{\Psi_{0}(t}).
\end{equation}
Using perturbation theory, the Hamiltonian is partitioned into a zeroth-order component, which describes the molecule in the absence of any time-dependent field, and a first-order component, which is the semi-classical interaction between the molecule and an external dynamic field,
\begin{equation}\label{H_one}
    \hat{H}(t) = \hat{H}^{(0)}+\hat{H}^{(1)}(t).
\end{equation}
There are two principal formalisms for clalculating response properties. The first involves the expansion of the time-dependent wavefunction and corresponding expectation-value properties, such as the electric dipole moment, in orders of the perturbation, followed by Fourier transformation to the frequency domain, yielding order-by-order property tensors such as the polarizability, optical activity tensor, etc.\cite{koch1990coupled} (See Appendix~\ref{Exp_value} for details.) The second approach identifies response functions as derivatives of the time-averaged quasi-energy with respect to external field strength parameters~\cite{SasaganeQuasi, christiansen1998response}. We make use of the latter formalism in this work. The quasi-energy formalism was first introduced by Sasagane~\cite{SasaganeQuasi} and later refined by H\"{a}ttig, Christiansen and  J\"{o}rgensen~\cite{christiansen1998response}.

 We can express the first-order perturbation component of the Hamiltonian in Eq.~\eqref{H_one} as a discrete sum of periodic perturbations as
\begin{equation}\label{Sum}
    \hat{H}^{(1)}(t)=\sum_{j=-N}^{j=N} e^{-\mathrm{i} \omega_j t} \hat{H}^{(1)}(\omega_j), \hspace{0.05 in} \hat{H}^{(1)}(\omega_j) = \sum_{Y} \epsilon_Y (\omega_j) \hat{Y},
\end{equation}
where $\hat{Y}$ is a frequency-independent operator describing the interaction between the external field and the molecular system and $\epsilon_Y$ is the frequency-dependent strength parameter associated with the given external field (see Ref.~\citenum{christiansen1998response}), while $N$ refers to the 
total number of monochromatic periodic perturbations. For example, $\hat{Y}$ corresponds to the dipole moment operator $(\vec{\mu})$ when the perturbation is an oscillating electric field and is associated with the magnetic moment operator $(\vec{m})$ in the case of an external magnetic field. This can be expressed in the second quantized formalism as 
\begin{equation}
    \hat{Y} = Y^{p}_{q} a^{\dagger}_p a_q,
\end{equation}
where $Y^{p}_{q}$ refers to $\langle \chi_p | \vec{\mu}_i | \chi_q \rangle$ with $i \in \{x,y,z\}$ in the case of an external electric field and $\langle \chi_p | \vec{m}_i | \chi_q \rangle$ for a magnetic field. The indices $p$, $q$ denote the molecular orbitals and the operators $a^{\dagger}_p$, $a_q$ are the usual fermionic creation and annihilation operators. It should be noted that $\vec{\mu}_i = -\vec{r}_i$ and $\vec{m}_i = -\frac{1}{2}(\vec{r}\times\vec{p})_i$, where $\vec{r}$ and $\vec{p}$ refer to the position and momentum vectors, respectively. Thus, the summation over $\hat{Y}$ in Eq.~\eqref{Sum} covers all the possible interactions of the molecular system with a given external field.
% and $\frac{1}{2} \langle \chi_p | \vec{r} \times \vec{p} | \chi_q \rangle$ in the case of external electric and magnetic fields, respectively, where $\vec{r}$ and $\vec{p}$ refer to the position and momentum vectors, respectively, and the indices $p$, $q$ denote the molecular orbitals.
To ensure that $\hat{H}^{(1)}(t)$ stays Hermitian, the operator $\hat{Y}$ should be Hermitian as well, along with other necessary conditions such as, $\omega_{-j} = -\omega_{j}$ and $\epsilon^{*}_Y (\omega_j) = \epsilon_Y(-\omega_j)$.

 The central quantity in this quasi-energy formalism is the time-dependent quasi-energy defined as
\begin{equation}
    Q(t)=\bra{\Psi_{0}(t)}\left(\hat{H}^{(0)}+\hat{H}^{(1)}(t) -\mathrm{i} \frac{\mathrm{d}}{\mathrm{d} t}\right)\ket{\Psi_{0}(t)}.
\end{equation}
The quasi-energy can be seen as an analogue of energy in the time-dependent domain. By invoking
the time-averaged time-dependent Hellmann-Feynman theorem~\cite{christiansen1998response}, one can obtain response functions by taking the derivatives of the time-averaged quasi-energy with respect to external field strength parameters.\\
In order to derive the response equations, we consider the following time-dependent ansatz of the wavefunction in the presence of an external field,
\begin{equation}
    \ket{\Psi_{0}(t)} = e^{\hat{R}(t)}\ket{\Psi_{0}},
\end{equation}
where the $\hat{R}(t)$ is linear cluster operator of the following form
% \begin{align}\label{R_OK}
%         \hat{R}(t)=\sum_i\sum_{\mu_{i}}\big[A_{\mu_i}(t) \hat{G}_{\mu_{i}}- A^{*}_{\mu_i^{\dagger}}(t)\hat{G}_{\mu_i^\dagger}\big].
% \end{align}
\begin{align}\label{R_resp}
    \hat{R}(t) & = \hat{R}_1(t) + \hat{R}_2(t) + \hat{R}_3(t) + \dots .
\end{align}   
The ground state $\ket{\Psi_\text{0}}$ here is the optimized ground state wavefunction obtained by a VQE algorithm on a quantum computer. %$\ket{\Psi_{\text{0}}} = \text{U}(\theta)\ket{0}$. 
We define the operators $\hat{R}_i$ ($i \in \{1,2,3,..\}$) using second-quantized excitation and de-excitation operators of $i^{\text{th}}$ rank as
\begin{align}\label{R_OK}
        \hat{R}_i(t)=\sum_{\mu}\big[A_{\mu_{i}}(t) \hat{G}_{\mu_{i}}+A^{*}_{\mu_i^{\dagger}}(t)\hat{G}_{\mu_i^\dagger}\big],
\end{align}
where $\hat{G}_{\mu_{i}}$ and $\hat{G}_{\mu^\dagger_{i}}$ refer to an excitation and de-excitation operator of rank $i$ with the corresponding response amplitudes  $A_{\mu_{i}}$(t) and $A^{*}_{\mu_i^{\dagger}}$(t), respectively.  
The value of  $i$ can, of course, range from 1 to $N$, where $N$ is the number of electrons in the system. The action of these operators on the reference wavefunction $|0\rangle$ --- Hartree-Fock (HF) in our case --- can be illustrated mathematically as
\begin{align}
\hat{G}_{\mu_{i}}|0\rangle =& \hspace{0.02 in}|\mu_i\rangle, \nonumber \\
\langle 0 | \hat{G}_{\mu_{i}^{\dagger}} =& \hspace{0.02 in}\langle \mu_i|,
\end{align}
where $|\mu_i\rangle$ denotes an ``excited'' Slater determinant of rank $i$.
One can expand the Fourier components of these response amplitudes in successive orders of the perturbation, just like in equation~\eqref{order_expansion} in the Appendix.
It can be shown that solving the time-dependent Schr\"{o}dinger equation is equivalent to the variational minimization of the time-averaged quasi-energy which is defined as $\{L(t)\}_{T} = \frac{1}{T} \int_{0}^{T} \mathrm{~d}t\hspace{0.02 in} Q(t)$.~\cite{christiansen1998response} 
After expanding the quasi-energy in different orders of the perturbation ($L(t) = L^{(0)}(t) + L^{(1)}(t) + L^{(2)}(t) + ...$), the equations for solving frequency-dependent response amplitudes of different orders can be obtained through the following equations,
\begin{align}\label{TAQE}
%&\{L(t)\}_{T} = \frac{1}{T} \int_{0}^{T} \mathrm{~d}t\hspace{0.02 in} Q(t),\nonumber\\
&\frac{\partial}{\partial A^{(m)}_{\mu_i}(\omega_j)} {\{L^{(n)}(t)\}_{T}} = 0,\\\nonumber
&\frac{\partial}{\partial A^{(m)*}_{\mu_i^{\dagger}}(\omega_j)} {\{L^{(n)}(t)\}_{T}} = 0,
\end{align}
where $m \leq n$. 
It should be noted that the response amplitudes satisfy the $2n+1$ rule, which states that for calculating a molecular property of perturbation order $2n+1$, one needs only up to order $n$ wavefunction parameters. 
Thus, first-order response amplitudes can provide up to third-order properties such as hyperpolarizabilities. 
Putting $m=1, n=2$ in Eq.~\eqref{TAQE}, one obtains the following secular equation for first-order response amplitudes associated with the perturbation operator $\hat{Y}$ at frequency $\omega_j$
\begin{align}\label{response_secular}
\bigg[\left(\begin{array}{cc}
\mathbf{M} & \mathbf{Q} \\
\mathbf{Q}^{*} & \mathbf{M}^{*}
\end{array}\right)
-\omega_{j}\left(\begin{array}{cc}
\mathbf{V} & \mathbf{W} \\
-\mathbf{W}^{*} & -\mathbf{V}^{*}
\end{array}\right)\bigg]
\left[\begin{array}{l}
\mathbf{A}^{(1)}_{Y}(\omega_j) \\
\mathbf{B}^{(1)}_{Y}(\omega_j)
\end{array}\right]\\
= \left[\begin{array}{l}
\begin{aligned}
\mathbf{Z}_{Y}& \\
\mathbf{-Z}^{*}_{Y}&
\end{aligned}
\end{array}\right],\nonumber
\end{align}
%where $(\mathbf{B}^{(1)}_{Y})_{\mu_i} = (\mathbf{A}^{*(1)}_{Y})_{\mu_i^\dagger}$
where $\mathbf{B}^{(1)}_{Y} = (\mathbf{A}^{(1)}_{Y})^{\dagger}$
and the elements of matrices $\textbf{M}$, $\textbf{Q}$, $\textbf{V}$, $\textbf{W}$ and vector $\mathbf{G_{Y}}$ are defined as
\begin{align}\label{MVQW}
    \text{M}_{\mu_i,\nu_j}= \hspace{0.02 in}& \langle \Psi_{\text{0}}|[\hat{G}_{\mu^{\dagger}_{i}},[\hat{\text{H}},\hat{G}_{\nu_{j}}]]|\Psi_{\text{0}}\rangle,   \\
    \text{V}_{\mu_i,\nu_j}= \hspace{0.02 in}& \langle \Psi_{\text{0}}|[\hat{G}_{\mu^{\dagger}_{i}},\hat{G}_{\nu_{j}}]|\Psi_{\text{0}}\rangle, \nonumber\\
    \text{Q}_{\hspace{0.01 in}\mu_i,\nu_j}= & -\langle \Psi_{\text{0}}|[\hat{G}_{\mu^{\dagger}_{i}},[\hat{\text{H}},\hat{G}_{\nu^{\dagger}_{j}}]]|\Psi_{\text{0}}\rangle, \nonumber \\
    \text{W}_{\mu_i,\nu_j}= & -\langle \Psi_{\text{0}}|[\hat{G}_{\mu^{\dagger}_{i}},\hat{G}_{\nu^{\dagger}_{j}}]|\Psi_{\text{0}}\rangle \nonumber \\
    Z_{Y}({\mu_i})= & \bra{ \Psi_{0}}[\hat{\text{Y}},\hat{G}_{\mu_i}]\ket{\Psi_{0}}\nonumber.
\end{align}

%\textcolor{red}{X and Y are not deefined anywhere. They should have some operator form?}
%{\color {red} and the elements of are defined as, define all the matrices as well.. shift from EOM-CC section!}
Finally, the response functions can be obtained by taking the derivative of the time-averaged quasi-energy of an appropriate order with respect to field strengths. For example, the linear response function can be obtained as,
\begin{align}\label{resp_function}
%&\langle X\rangle=\frac{d\{Q\}_{T}}{d \varepsilon_{x}(0)} \nonumber\\
\langle\langle X ; Y\rangle\rangle_{\omega_j}&=\frac{\partial ^{2}{\{L^{(2)}(t)\}_{T}}}{\partial \varepsilon_{X}\left(-\omega_{j}\right) \partial \varepsilon_{Y}\left(\omega_j\right)}\\
\vspace{0.5 in}
&= \mathbf{Z_{X}}\boldsymbol{\cdot}\mathbf{A}_{Y}(\omega_j) +
    \mathbf{Z^{*}_{X}}\boldsymbol{\cdot}\mathbf{B}_{Y}(\omega_j),\nonumber
% &\langle\langle X ; Y, Z\rangle\rangle_{\omega_j, \omega_{k}}=\frac{\partial^{3}{\{L^{(3)}(t)\}_{T}}}{\partial \varepsilon_{X}\left(-\omega_j - \omega_k \right)\partial \varepsilon_{Y}\left(\omega_j\right) d \varepsilon_{Z}\left(\omega_k\right)}.\nonumber
\end{align}
%{\color{red} mention about sum of frequencies as zero!}. 
% From equation~\ref{resp_function}, the linear response function takes the following form,
% \begin{equation}
%     \langle\langle X;Y\rangle\rangle_{\omega_{Y}}= \mathbf{G_{X}}\boldsymbol{\cdot}\mathbf{A}_{Y}(\omega_Y) +
%     \mathbf{G^{*}_{X}}\boldsymbol{\cdot}\mathbf{B}_{Y}(\omega_Y),
% \end{equation}
where $Z_{X}({\mu_i})=  \bra{ \Psi_{0}}[\hat{\text{X}},\hat{G}_{\mu_i}]\ket{\Psi_{0}}$ and $\boldsymbol{\cdot}$ refers to the dot product operation. 
For exact electronic states, the linear response function can also be
written as a \texttt{SoS} expression~\cite{koch1990coupled},
\begin{equation}\label{SoS}
\begin{aligned}
\langle\langle X;Y\rangle\rangle_{\omega_{j}}=& \sum_{k>0} \frac{\bra{ \Psi_{\text{0}}}\hat{X}\ket{\Psi_{\text{k}}}\bra{\Psi_{\text{k}}}\hat{Y}\ket{\Psi_{0}}}{\omega_{j}-\omega_{\text{k}}} \\
&-\sum_{k>0} \frac{\bra{ \Psi_{\text{0}}}\hat{Y}\ket{\Psi_{\text{k}}}\bra{\Psi_{\text{k}}}\hat{X}\ket{\Psi_{0}}}{\omega_{j}+\omega_{\text{k}}},
\end{aligned}
\end{equation}
where $\bra{\Psi_{\text{k}}}$ refers to the wavefunction of the $k^\text{th}$ excited state with the excitation energy of $\omega_k$. Calculation of properties like specific rotation using the \texttt{SoS} formalism can be
computationally prohibitive as thousands of electronic excited states may need be evaluated to ensure convergence of eq.~\eqref{SoS}\cite{wiberg2006sum}. However, the \texttt{SoS} approach has its own advantages as well, specially for resonant- and near-resonant responses, where one just needs only excited states within a desired spectral window.
% It can be seen that the above equation involves a summation over all the excited states, which makes the \texttt{SoS} approach computationally prohibitive.
The linear response approach avoids the explicit calculation of all excited states by parametrizing the perturbation of the ground-state wavefunction in the presence of an external field through response amplitudes, which are solved through a linear system of equations. 
Furthermore, one can also get the values of excitation energies and transition moments for a given excited state by identifying the poles and evaluating the residues of the linear response function at poles, respectively~\cite{koch1990coupled}. It should be noted that the values of excitation energies (EEs), ionization potentials (IPs), and electronic affinities (EAs) calculated using the \texttt{qLR} approach should be identical to the ones obtained from the \texttt{qEOM} approach~\cite{koch1990coupled}. Please refer to the Appendix (section~\ref{EOM}) for a detailed theoretical background of the \texttt{qEOM} method. 
In an earlier work~\cite{asthana2022equation}, we have shown that the \texttt{qEOM} method does not necessarily satisfy the ``killer condition'', leading to large errors for IPs and EAs even for small molecular systems. Thus, one needs to make sure that the \texttt{qLR} approach also compiles with the ``killer condition'' in order to obtain accurate molecular response properties.
%The $i,j$ element of the dipole polarizability and optical rotation tensors can be written as
%$\langle\langle \hat{\boldsymbol{\mu}}_i , \hat{\boldsymbol{\mu}}_j\rangle\rangle_\omega$ and $\langle\langle \hat{\boldsymbol{\mu}}_i, \mathbf{\hat{L}}_j\rangle\rangle_\omega$, respectively.
\subsection{Vacuum annihilation or ``Killer'' condition}\label{VAC}
The vacuum annihilation condition (VAC) states that ground state cannot be de-excited since it is the lowest energy eigenstate, i.e., 
\begin{equation}
\hat{\mathbb{O}}^\dagger_\text{k}\ket{\Psi_{\text{0}}} = 0,
\end{equation}
where $\hat{\mathbb{O}}_\text{k}$ is a state-transfer operator such that its action on the ground state leads to the $k^{\text{th}}$ excited state,
\begin{align}
    \hat{\mathbb{O}}_\text{k}|\Psi_0\rangle=\ket{\Psi_k}.
\end{align}
%These conditions are related to the resolution of identity in the operator space for a given vacuum state (ground-state wavefunction)~\cite{prasad1985some}.
It is easy to see that the VAC is satisfied for an exact state-transfer operator~\cite{szekeres2001killer,prasad1985some,mertins1996algebraic,weiner1980calculation,hodecker2020unitary,levchenko2004equation} by writing it in a projector form as
\begin{align}
     \hat{\mathbb{O}}_{\text{k}} & = \ket{\Psi_{\text{k}}}\bra{\Psi_{\text{0}}}.
 \end{align}
 The application of the adjoint of the exact state-transfer operator on the ground-state wavefunction produces null, i.e, 
 \begin{align}
 \hat{\mathbb{O}}^\dagger_\text{k}|\Psi_{\text{0}}\rangle&=\ket{\Psi_{\text{0}}}\bra{\Psi_{\text{k}}}\ket{\Psi_{\text{0}}}=0\quad \forall\hspace{0.02 in} \text{k}, \label{VACeq}
\end{align}
since the ground and excited state wavefunctions are always orthogonal to each other. However, 
the VAC may not be satisfied for approximate state-transfer operators. For instance, the VAC is not necessarily satisfied for a general state-transfer operator defined in the \texttt{qEOM} formalism (see Eq.~\eqref{OK}), i.e.
\begin{equation}
\hat{\mathbb{O}}^\dagger_\text{k}|\Psi_{\text{0}}\rangle=
 \sum_i\sum_{\mu_{i}}\big[(A^\text{k})_{\mu_i^\dagger}^{*} \hat{G}_{\mu_i^\dagger}+(B^{\text{k}})_{\mu_{i}}^{*}\hat{G}_{\mu_i}\big]
 |\Psi_{\text{0}}\rangle \neq 0.\label{VAC_neq_0}
\end{equation}
This can lead to non-orthogonal ground and excited state wavefunctions and produce large errors in charged excitation energies~\cite{asthana2022equation}.% for even small molecular systems.
%Satisfaction of killer condition is also necessary for appropriate resolution of identity in a truncated operator manifolds, as discussed in Ref.\cite{prasad1985some}.
Two distinct methods were proposed to satisfy the VAC
%There are two successful ways of satisfying the VAC 
for approximate state-transfer operators, namely, self-consistent operators~\cite{prasad1985some}
%(discussed in ~\ref{sc}) 
and projection based approaches~\cite{szekeres2001killer}
%(discussed in ~\ref{proj}) 
(both discussed below) leading to two different formalisms for estimating excited-state properties.

\subsubsection{Self-consistent operators}\label{sc}
One way to ensure that the VAC is always satisfied is through the use of a self-consistent operator manifold instead of the manifold defined using HF as the reference.
This approach was originally introduced by Prasad and Mukherjee~\cite{prasad1985some} for methods with unitary parametrization of ground-state wavefunction. 
The self-consistent manifold  can be defined using the primitive excitation manifold ($\hat{G}_{\mu_i}\cup \hat{G}_{\mu_i}^\dagger$) as 
\begin{align}
    \hat{S}_{\mu_i} &= \text{U}(\theta)\hspace{0.02 in}\hat{G}_{\mu_i}\hspace{0.01 in}\text{U}^{\dagger}(\theta),
\end{align}
where $\text{U}(\theta)$ refers to the unitary operator used to obtain the ground-state wavefunction ($\ket{\Psi_{\text{0}}}$).

It can be seen that the application of the action of the adjoint of a general state transfer operator, defined using the operators from the self-consistent operator manifold, on the ground-state wavefunction is zero, i.e.,
\begin{equation}\label{deexc_sc}
\begin{split}
     \mathbb{O}^{\dagger}_k\ket{\Psi_{\text{0}}} &= \sum_{\mu_i}(A^\text{k})_{\mu_i^\dagger}^{*}\text{U}(\theta)\hat{G}_{\mu_i}^\dagger\text{U}^{\dagger}(\theta)\text{U}(\theta)\ket{0}\\  &= \sum_{\mu_i}(A^\text{k})_{\mu_i^\dagger}^{*}\text{U}(\theta)\hat{G}_{\mu_i}^\dagger \ket{0} = 0,
\end{split}
\end{equation}
as the regular de-excitation operator acting on the reference wavefunction yields zero. It should be noted that this formalism is general and applies to any wavefunction ansatz where the ground state is obtained through an action of a unitary operator acting on a starting state such as HF. Similar approaches have been developed for excited-state methods using unitary coupled cluster (UCC) theory~\cite{liu2018unitary}. 

Use of the self-consistent operator manifold in \texttt{qEOM} gives rise to the following simplified working equation
\begin{align}\label{sc_eig}
    \mathbf{M^{\text{sc}}}\mathbf{A}_k=\text{E}_{\text{0k}}\mathbf{A}_k.
\end{align}
where,
\begin{align}\label{sceqM}
\text{M}^{\text{sc}}_{\mu_i,\nu_j}=&\hspace{0.02 in}\langle \Psi_{0}|\hat{G}_{\mu_i^\dagger}\text{U}^\dagger(\theta)\hat{\text{H}}\text{U}(\theta)\hat{G}_{\nu_j}|\Psi_{0}\rangle-\delta_{\mu_i,\nu_j}*\text{E}_{\text{0}}.
\end{align}
For more details, please refer to 
Ref.~\cite{asthana2022equation}.
%Compared with the original secular equation associated with QSE and qEOM, this formalism doesn't require solving generalized eigenvalue equation. Generalized eigenvalue equation for excited state methods has been shown to be ill-conditioned and sensitive to noise~\cite{asthana2022equation}.
%To summarize, our formalism satisfies the VAC and provides size-intensive energy differences. It also compares favorably with the classical EOM method which suffers from a number of drawbacks~\cite{liu2018unitary} due to the non-hermitian nature of the eigenvalue equation.

Using the self-consistent operator manifold, the response equations obtained in Eq.~\eqref{response_secular} are also simplified and can now be separated into two equations
\begin{align}\label{response_secular_new_sc}
\begin{array}{cc}
(\mathbf{M^{\text{sc}}} - \omega_Y \textbf{I}) \\ 
(\mathbf{M^{\text{sc}}} + \omega_Y \textbf{I})
\end{array}
\begin{array}{l}
\mathbf{A}_{Y} (\omega_Y) = \\
\mathbf{B}_{Y} (\omega_Y) = 
\end{array}
\begin{array}{l}
\begin{aligned}
& \mathbf{Z}^{\text{sc}}_{Y}, \\
& \mathbf{-Z}^{\text{sc}*}_{Y},
\end{aligned}
\end{array}
\end{align}
where $Z^{\text{sc}}_{Y}(\mu_i)= \bra{ \Psi_{\text{0}}}\text{U}^\dagger(\theta) \hat{\text{Y}}\text{U}(\theta)\hat{G}_{\mu_i}\ket{\Psi_{\text{0}}}$. 
However, one can combine the above two equations into one single equation, in order to lower the computational costs involved. For example, if we consider the perturbation to be electric-dipole based, $\mathbf{Z^{\text{sc}}_Y}$ is identical to $\mathbf{Z^{\text{sc}*}_Y}$ and we arrive at the following equation,
\begin{align}\label{response_secular_single}
&((\mathbf{M^{sc}})^{2} - \omega_Y^2\textbf{I}) (\mathbf{A_Y}(\omega_Y) - \mathbf{B_Y}(\omega_Y)) = 2* \mathbf{M^{\text{sc}}}\mathbf{Z^{\text{sc}}_Y},
%&\mathbf{A_Y} + \mathbf{B_Y} = \frac{\mathbf{M}}{\omega_Y}(\mathbf{A_Y} - \mathbf{B_Y},)\nonumber\\
% &\langle\langle \textbf{X}, \textbf{Y}\rangle\rangle_{\omega_{Y}}= \mathbf{G_{X}}\cdot(\mathbf{M}{\omega_Y}\mathbf{A}_{Y}(\omega_Y) -
%   \mathbf{B}_{Y}(\omega_Y)).\nonumber
\end{align}
and the linear response function can be reformulated as 
\begin{align}
&\langle\langle \textbf{X}, \textbf{Y}\rangle\rangle_{\omega_{Y}}= \frac{1}{\omega_Y}\mathbf{Z^{\text{sc}}_{X}}\boldsymbol{\cdot}(\mathbf{M^{\text{sc}}}(\mathbf{A}_{Y}(\omega_Y) -
    \mathbf{B}_{Y}(\omega_Y)).
%    \mathbf{B}_{Y}(\omega_Y)],$
\end{align}

\subsubsection{Projection operators} \label{proj}
Surj\a'an and co-workers developed the projection operator technique~\cite{szekeres2001killer} to ensure that the VAC is always satisfied while calculating molecular ionization potentials. The   projected excitation operator ($\hat{S}_{\mu_i}$) can be written as
\begin{equation}
      \hat{S}_{\mu_i} =  \hat{G}_{\mu_i} \ket{\Psi_{\text{0}}}\bra{\Psi_{\text{0}}}.
\end{equation}
For non-number-conserving operators (which appear in ionization potential or electron affinity  calculations), it can be easily seen that the action of the projected de-excitation operator on the ground-state wavefunction vanishes, i.e.,
\begin{align}\label{deexc_pro}
     \hat{S}_{\mu_i}^\dagger\ket{\Psi_{\text{0}}} &= \ket{\Psi_{\text{0}}}\bra{\Psi_{\text{0}}}\hat{G}_{\mu_i}^\dagger\ket{\Psi_{\text{0}}} = 0.
\end{align}
Fan and co-workers~\cite{fan2021equation,liu2022quantum} recently made use of these operator manifolds within the framework of equation of motion theory to calculate band structures on a quantum computer. To ensure that Eq.~\eqref{deexc_pro} also holds true for number-conserving operators, we shift all the operators by their expectation values 
\begin{equation}
    \hat{\bar{G}}_{\mu_i} = \hat{G}_{\mu_i} - \bra{\Psi_{\text{0}}}\hat{G}_{\mu_i}\ket{\Psi_{\text{0}}}.
\end{equation}
This can also be seen as a form of a normal ordering of the operators with respect to a general reference wavefunction~\cite{Kutzelnigg_MK_ENO}.
% The state-transfer operator written in terms of these shifted projected operators satisfy the VAC, which can be demonstrated as
% \begin{equation}
% \hat{\mathbb{O}}_\text{k}^\dagger\ket{\Psi_{\text{0}}} =
%  \sum_i\sum_{\mu_{i}}\big[(A^\text{k})_{\mu_i}^{*}\ket{\Psi_{\text{0}}} \cancelto{0}{\bra{\Psi_{\text{0}}}\hat{\bar{G}}_{\mu_i}^\dagger\ket{\Psi_{\text{0}}}}. \label{VAC_pro}
% \end{equation}
Just like the self-consistent formalism, this approach is quite general and can be used   with any wavefunction ansatz.
Using the shifted projected operators in Eq.~\eqref{secular} (see Appendix~\ref{EOM}), one gets a generalized Hermitian eigenvalue equation
\begin{align}\label{pro_eig}
    \mathbf{M^{\text{proj}}}\mathbf{A}_k=\text{E}_{\text{0k}}\mathbf{V^{\text{proj}}}\mathbf{A}_k,
\end{align}
where
\begin{align}
\text{M}^{\text{proj}}_{\mu_i,\nu_j}=& \bra{\Psi_{\text{0}}} \hat{\bar{G}}_{\mu_i}^\dagger
\hat{\text{H}}\hspace{0.02 in} \hat{\bar{G}}_{\nu_j}
\ket{\Psi_{\text{0}}} ,\\
\text{V}^{\text{proj}}_{\mu_i,\nu_j}=&\bra{\Psi_{\text{0}}} \hat{\bar{G}}_{\mu_i}^\dagger \hat{\bar{G}}_{\nu_j}
\ket{\Psi_{\text{0}}}\nonumber.
\end{align}
Eq.~\eqref{pro_eig} looks very similar to the one obtained in the QSE approach, except that the identity operator is not involved in the operator pool. Stated differently, unlike the QSE approach, the ground-state wavefunction does not participate in the diagonalization procedure, which ensures size-intensive excitation energies. However, the evaluation of the overlap matrix makes it more susceptible to noise~\cite{asthana2022equation}, compared   to the self-consistent operator approach. 

The equations for calculating the response amplitudes is simplified as well when we make use of these shifted projection operators.  Eq.~\eqref{MVQW} now can be decoupled into 
two separate equations,
%(\textcolor{red}{M V and G have different expressions for qLR(sc) and proj. They should have a prime or something to distinuish them.})
\begin{align}\label{response_secular_new_pro}
\begin{array}{cc}
(\mathbf{M}^{\text{proj}} - \omega_Y \textbf{V}^{\text{proj}}) \\ 
(\mathbf{M}^{\text{proj}} + \omega_Y \textbf{V}^{\text{proj}})
\end{array}
\begin{array}{l}
\mathbf{A}_{Y} (\omega_Y) = \\
\mathbf{B}_{Y} (\omega_Y) = 
\end{array}
\begin{array}{l}
\begin{aligned}
\mathbf{Z}^{\text{proj}}_{Y}& \\
\mathbf{-Z}^{\text{proj}*}_{Y}&
\end{aligned}
\end{array},
\end{align}
where  $Z^{\text{proj}}_{Y}(\mu_i)= \bra{ \Psi_{\text{0}}}\hat{\text{Y}}\hat{G}_{\mu_i}\ket{\Psi_{\text{0}}}$. Of course, one can combine the two equations into a single one and obtain equations similar to the self-consistent approach [Eq.~\eqref{response_secular_single}] with the identity matrix $\mathbf{I}$ replaced by the overlap matrix $\mathbf{V}^{\text{proj}}$.\\
\section{Proposed implementation}\label{implementation}

Here, we discuss the proposed implementation of \texttt{qLR(sc)} and \texttt{qLR(proj)} methods on near-term quantum computers. The working equations of \texttt{qLR(sc)} method are given in Eq.~\eqref{response_secular_single}, which requires the evaluation of matrices $\mathbf{M}^{\text{sc}}$ and $\mathbf{Z^{\text{sc}}_Y}$ on a quantum computer, after which the resulting equation is solved classically. 
The creation of the matrix $\mathbf{M}^{\text{sc}}$ requires the creation of diagonal and off-diagonal terms defined by Eq.~\eqref{sceqM}.
The evaluation of matrix $\mathbf{M}^{\text{sc}}$ can be carried out by the methods discussed in Ref.~\cite{asthana2022equation} without the use of any ancilla qubits.
To summarize, the diagonal elements can be evaluated as expectation value of $\hat{H}$ using the pre-optimized circuit obtained during the VQE procedure for estimating the ground state wavefunction. However, instead of the HF state, singly and doubly excited Slater determinants are now used as the reference, e.g.,
\begin{align}
    M^{\text{sc}}_{\mu_i,\mu_i}&=\bra{0}\hat{G}^\dagger_{\mu_i}U^\dagger(\theta)\hat{H}U(\theta)\hat{G}_{\mu_i}\ket{0}.
\end{align}
The off-diagonal elements, for which popular algorithms use the Hadamard test for evaluation, can be evaluated in a much simpler fashion using the relationship
\begin{align}\label{offdiagonalexp}
   \hbox{Re}[M_{\mu_i,\mu_j}]=M_{\mu_i+\mu_j,\mu_i+\mu_j}-\frac{M_{\mu_i,\mu_i}}{2}-\frac{M_{\mu_j,\mu_j}}{2},
\end{align}
where the term $M_{\mu_i+\mu_j,\mu_i+\mu_j}$ is given by
\begin{align}\label{mijexp}
\begin{split}
&M_{\mu_i+\mu_j,\mu_i+\mu_j}=\\&\bra{0} \frac{1}{\sqrt 2}(\hat{G}_{\mu_i}+\hat{G}_{\mu_j})^\dagger U(\theta)^\dagger \hat{H}U(\theta)\frac{1}{\sqrt 2}(\hat{G}_{\mu_i}+\hat{G}_{\mu_j})\ket{0}.
\end{split}
\end{align}
The creation of entanglement $(\hat{G}_{\mu_i}+\hat{G}_{\mu_j})\ket{0}$ can be simply achieved by using a Hadamard gate along with a few CNOTs (maximum of seven CNOTs required).

\begin{figure*}[tbp]
%\captionsetup[subfigure]{justification=centering,font=scriptsize,labelfont=large}
    %\centering
    \hspace*{\fill}%
    \subfloat{
        \includegraphics[width = 0.25\textwidth,height=4cm]{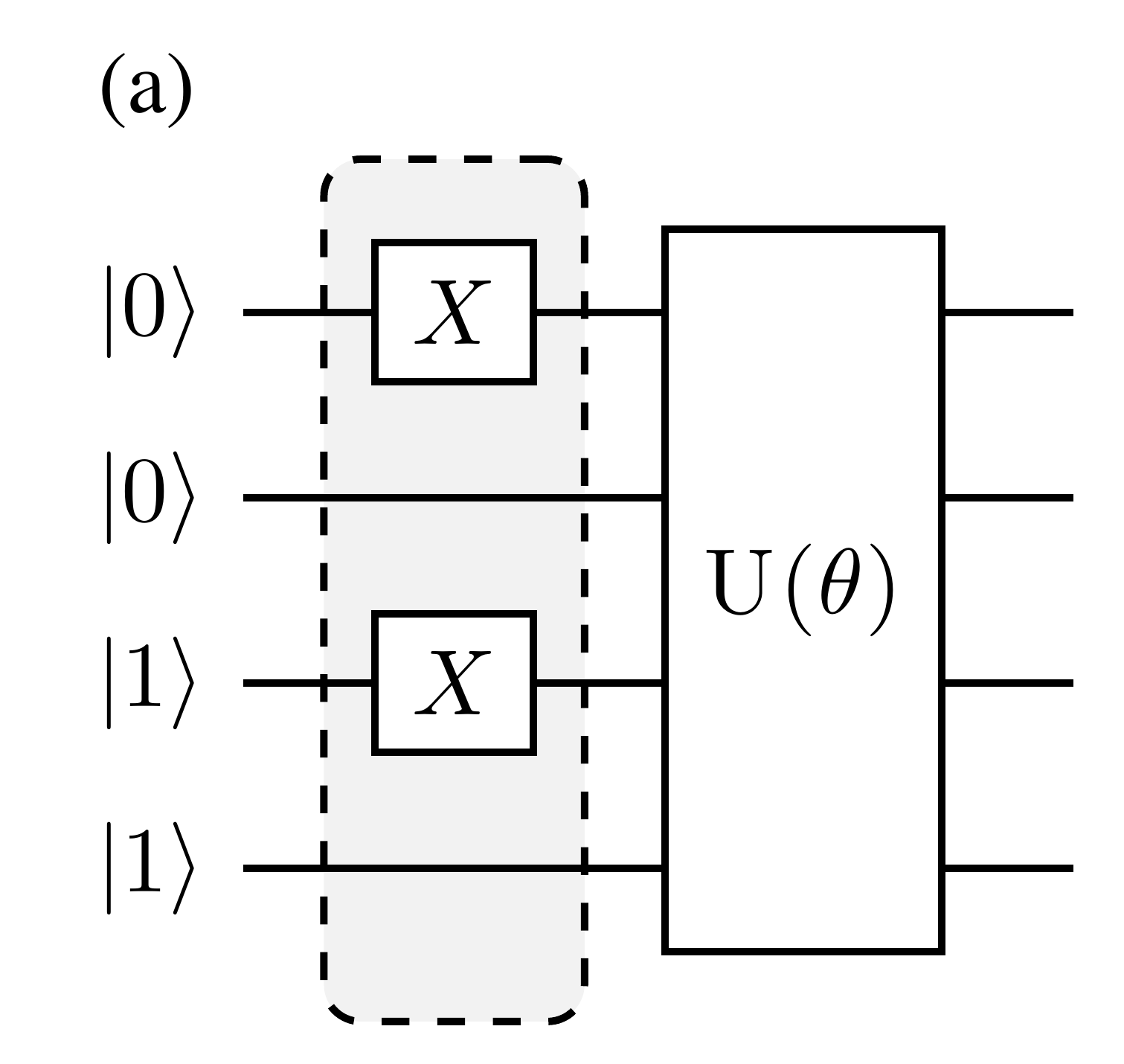}\label{circuitd}
    }\hfill
    \subfloat{
    \includegraphics[width = 0.4 \textwidth,height=4cm]{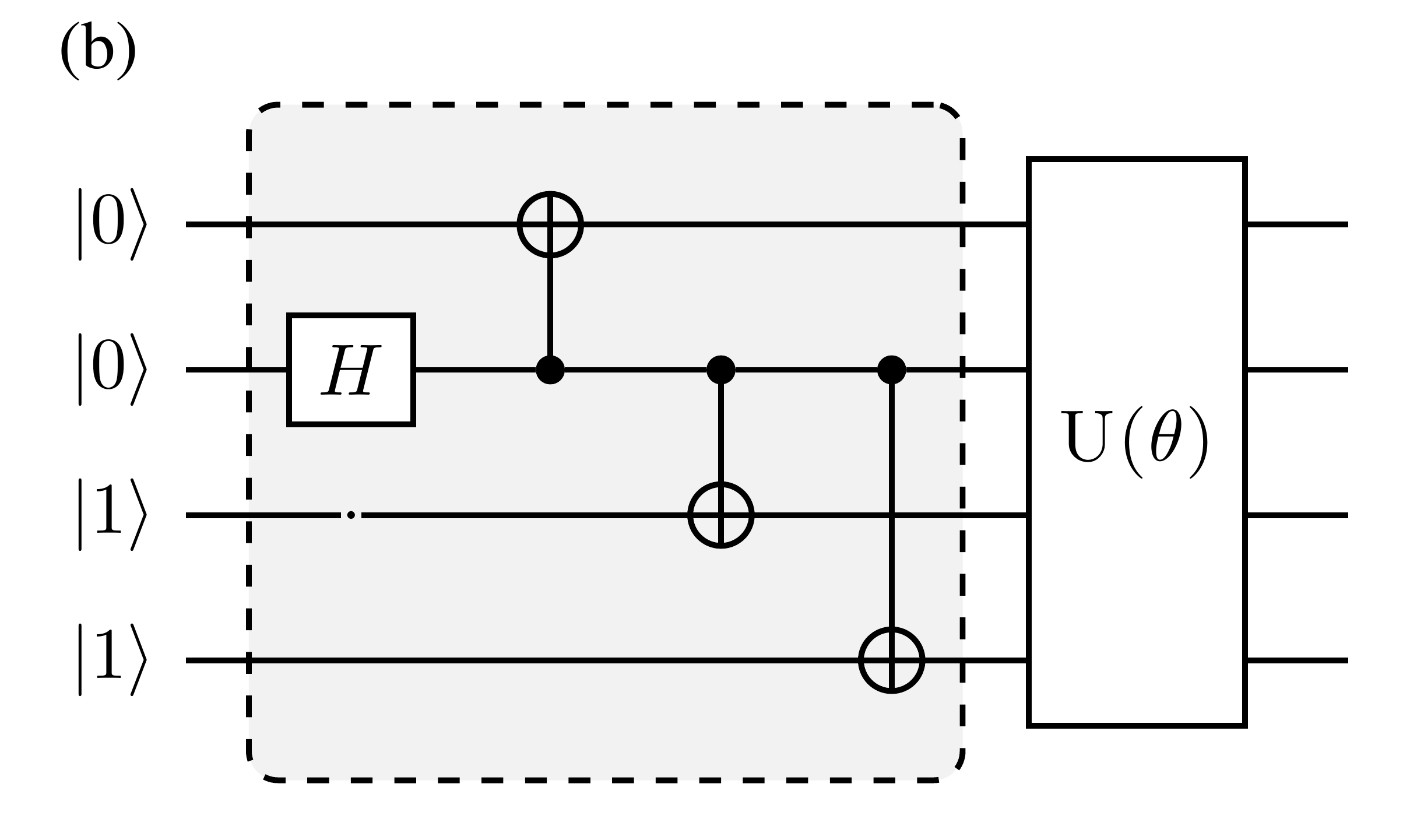}\label{circuitoff}
    }
    \caption{Proposed quantum circuit for the estimation of a representative element of the $\mathbf{Z_Y^\prime}$ vector element for using (a) an excited Slater determinant as the reference state and (b) an entangled state involving the HF state and excited Slater determinants as the reference state.}
    %\label{circuit}
    \hspace*{\fill}%
\end{figure*}

The matrix elements of vector $\mathbf{Z_Y}$ can be similarly computed. 
All elements of this matrix are analogous to the off-diagonal elements of matrix $\mathbf{M}$. 
It can be computed using the relationship
\begin{align}
    Z^{\text{sc}}_{Y}({\mu_i})= Z^{\prime}_{Y}({\text{0}+\mu_i})-\frac{Z^{\prime}_{Y}({\mu_i})}{2}-\frac{Z^{\prime}_{Y}({\text{0}})}{2},
\end{align}
where 
\begin{align}\begin{split}
     Z^{\prime}_{ Y}({\text{0}+\mu_i})= & \bra{ 0}(\hat{I}+\hat{G}_{\mu_i})^\dagger\text{U}^\dagger(\theta) \hat{\text{Y}}\text{U}(\theta)(\hat{I}+\hat{G}_{\mu_i})\ket{0}\\
     Z^{\prime}_{ Y}({\mu_i})= & \bra{ 0}\hat{G}_{\mu_i}^\dagger\text{U}^\dagger(\theta) \hat{\text{Y}}\text{U}(\theta)\hat{G}_{\mu_i}\ket{0}\\
     Z^{\prime}_{ Y}({\text{0}})= & \bra{0}\text{U}^\dagger(\theta) \hat{\text{Y}}\text{U}(\theta)\ket{0}.
     \end{split}
\end{align}
The element $Z^{\prime}_{ Y}({\text{0}})$ can be computed once using the ground state circuit, while the other two elements of $Z^{\prime}_{ Y}$ needed for $Z^{Y}({\mu_i})$ can be evaluated separately for each element of the $\mathbf{Z_{Y}}$ vector.
The elements $Z^{\prime}_{ Y}({\mu_i})$ and $Z^{\prime}_{ Y}({\text{0}})$ can be computed by measuring the expectation value of operator $\hat{Y}$ using states $\ket{\Psi_{\mu_i}}$ (see Fig.~\ref{circuitd} for an example) and $\ket{\Psi_{\text{0}}}$, respectively, where 
$\ket{\Psi_X}=U(\theta)\hat{G}_X\ket{0}$.
The element $Z^{\prime}_{ Y}({\text{0}+\mu_i})$ can be evaluated by measuring expectation value of operator $\hat{Y}$ using state $\ket{\Psi_{\text{0}+\mu_i}}$, which is prepared using a superposition of states  $\ket{\text{0}}$ and $\ket{\Psi_{\mu_i}}$ (see Fig. \ref{circuitoff} for an example).

%\begin{center}
%\begin{quantikz}[ row sep={20pt,between origins}]
%\lstick{$\ket{0}$}& \gate{X}\gategroup[4,steps=1,style={dashed,rounded corners,fill=gray!10, inner xsep=2pt}, background]{{}} & \gate[4]{\textsc{U($\theta$)}} & \qw \\
%\lstick{$\ket{0}$}& \qw & & \qw\\
%\lstick{$\ket{1}$}& \gate{X} & &\qw\\
%\lstick{$\ket{1}$}& \qw & &\qw 
%\end{quantikz},
%\end{center}

%\begin{center}
%\begin{quantikz}[ row sep={20pt,between origins}]
%\lstick{$\ket{0}$}& \gate{X}\gategroup[4,steps=5,style={dashed,rounded corners,fill=gray!10, inner xsep=2pt},background]{{}}   &\qw    & \targ{} & \qw &\qw   & \gate[4 ]{\textsc{U($\theta$)}} &  \qw  \\
%\lstick{$\ket{0}$}& \qw      & \gate{H}&\ctrl{-1} & \ctrl{1}   &\ctrl{2}&                            &\qw\\
%\lstick{$\ket{1}$}& \gate{X} & \qw.  & \qw  &  \targ{}    & \qw    &                          & \qw \\
%\lstick{$\ket{1}$}&\qw       &\qw    & \qw  &  \qw  &\targ{}  &                            &\qw 
%\end{quantikz},
%\end{center}

In the case of \texttt{qLR(proj)}, the matrices $\mathbf{M^{\text{proj}}}$, $\mathbf{V^{\text{proj}}}$ and the vector $\mathbf{Z_Y^{\text{proj}}}$ in Eq.~\eqref{response_secular_new_pro} can be computed using an estimate of the reduced density matrices (RDMs) using the prepared ground state. 
Evaluation of $\mathbf{M^{\text{proj}}}$, $\mathbf{V^{\text{proj}}}$ and $\mathbf{Z^Y}$ will involve the estimation of up to 6-, 4- and 3-body RDMs, respectively.
The scaling of the shot count is dominated by the estimation of the matrix $\mathbf{M}$ which scales as $O(N^{12})$ for both \texttt{qLR(sc)} and \texttt{qLR(proj)} approaches, where $N$ is some measure of the system size.
These requirements may be reduced by utilizing commutators which lead to the  cancellations of uncontracted terms~\cite{liu2021efficient}, approximations for higher-body RDMs and taking advantage of the high symmetry of the $\mathbf{M}$ matrix (such as by the use of Krylov-subspace-based algorithms like the Davidson method). The pathways to reduce computational complexity will be a topic of later studies. 

\section{Computational Details}\label{compdet}
All the calculations in this work employ the STO-3G basis set.
The second-quantized Hamiltonian is generated by the PySCF~\cite{Pyscf} software package and transformed into the Pauli representation using the Jordon-Wigner mapping function of the  OpenFermion~\cite{openfermion} program. 
We use a state-vector simulator to test the accuracy of the methods developed in this work.
The fermionic \texttt{ADAPT-VQE} method~\cite{grimsley2019adaptive} is employed to calculate the ground-state wavefunction using an operator pool composed of generalized singles and doubles excitation operators. 
Two classes of operator manifolds (self-consistent and projection operators that ensure that the VAC is always satisfied) are referred in shorthand notation as \texttt{sc} and \texttt{proj}.
Thus, the \texttt{qEOM} framework utilizing these operator manifolds are named as \texttt{q-sc-EOM} and \texttt{q-proj-EOM}. Similarly, we name the quantum formulation of linear response theory (\texttt{qLR}) using these operator manifolds as \texttt{qLR(sc)} and \texttt{qLR(proj)}. 
In this work, the state-specific properties like excitation energies, oscillator strengths, and rotational strengths are evaluated using the quantum equation of motion approaches (\texttt{q-sc-EOM}, \texttt{q-proj-EOM}) while dipole polarizabilities and specific rotation are calculated using the \texttt{qLR} theory (\texttt{qLR(sc)}, \texttt{qLR(proj)}). It should be noted that all these approaches utilize the ground-state wavefunction obtained from the \texttt{ADAPT-VQE} algorithm with gradient convergence criteria set to $10^{-3}$ $E_h$.
The code used for generating the data in this work can be found in Ref.~\cite{giteom}.

\section{Results}\label{results}
% just mention the formula for isotropic polarizability and  specific rotation! mention 1 or 2 lines before the results section!
\subsection{H$_2$}
% \begin{figure*}
%     \centering
%     \begin{minipage}{0.45\linewidth}
%         \centering
%         \includegraphics[width=0.9\textwidth]{h2_EE.pdf} % first figure itself
%         \caption{Excitation energies ($E_h$) of the H$_2$ molecule with the STO-3G basis set calculated using the \texttt{q-sc-EOM} and \texttt{q-proj-EOM} approaches as a function of the inter-hydrogen distance. The reference \texttt{FCI} values are plotted in grey and the deviations from the reference are shown in the upper panel, where the shaded region indicates errors below 0.1 eV. }
%         \label{fig:h2_EE}
%     \end{minipage}\hfill
%     \begin{minipage}{0.45\linewidth}
%         \centering
%         \includegraphics[width=0.9\textwidth]{h2_polar_error_combined.pdf} % second figure itself
%         \caption{Isotropic dynamic electric-dipole polarizability of the H$_2$ molecule with the STO-3G basis calculated at 589 nm using the \texttt{qLR(sc)} and \texttt{qLR(proj)} approaches as a function of the inter-hydrogen distance. The reference \texttt{SoS(FCI)} values are plotted in grey and the deviations from the  reference are shown in the upper panel, where the shaded region indicates errors below 1\%.}
%         \label{fig:h2_polar}
%     \end{minipage}
% \end{figure*}

\begin{figure*}[htp] 
    \centering
     %\captionsetup[subfigure]{justification=centering}
     \captionsetup{justification=raggedright,singlelinecheck=false}
    \subfloat{%
        \includegraphics[width=0.47\textwidth]{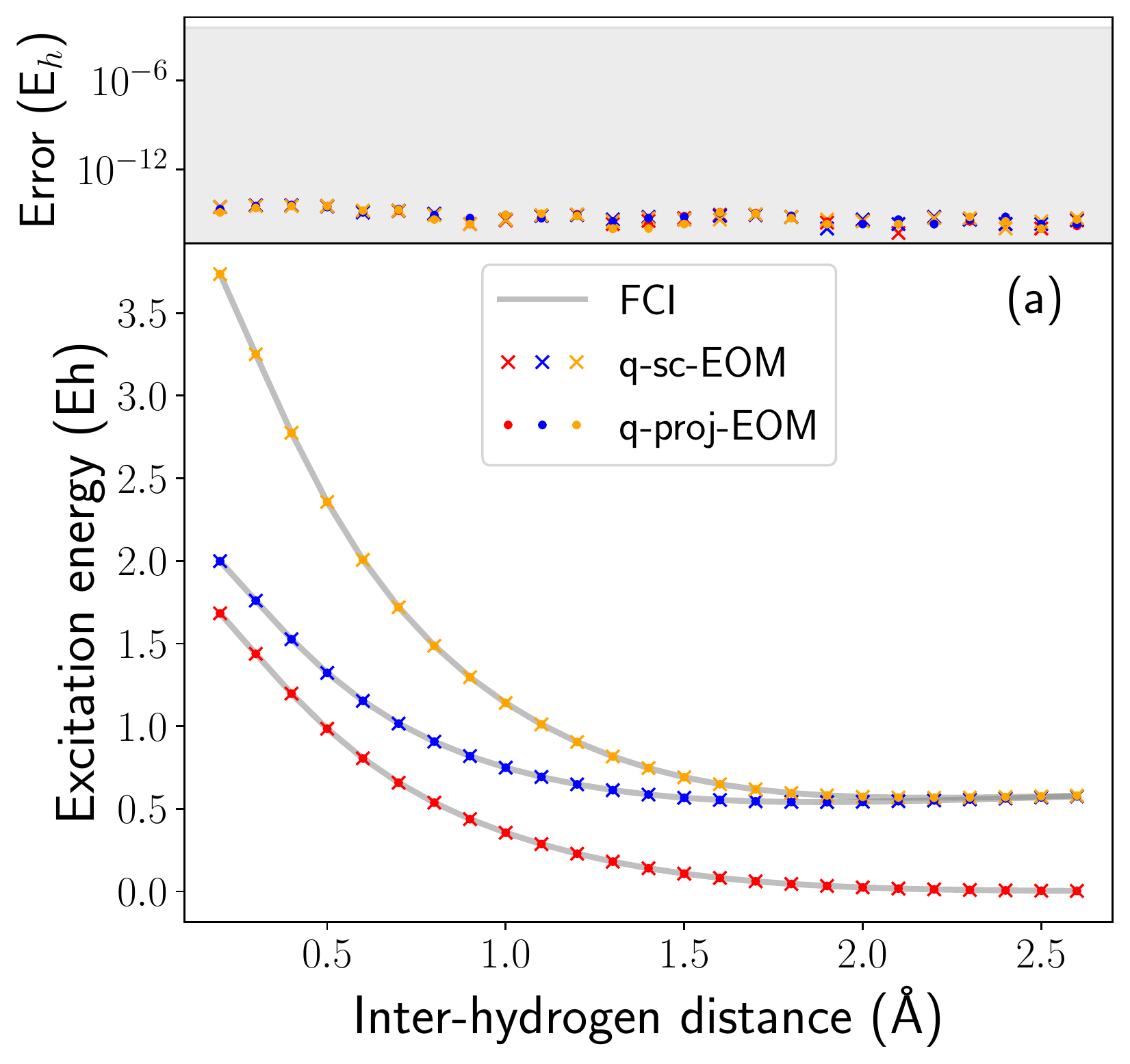}%
        \label{fig:h2_EE}%
        }%
    \hfill%
    \subfloat{%
        \includegraphics[width=0.47\textwidth]{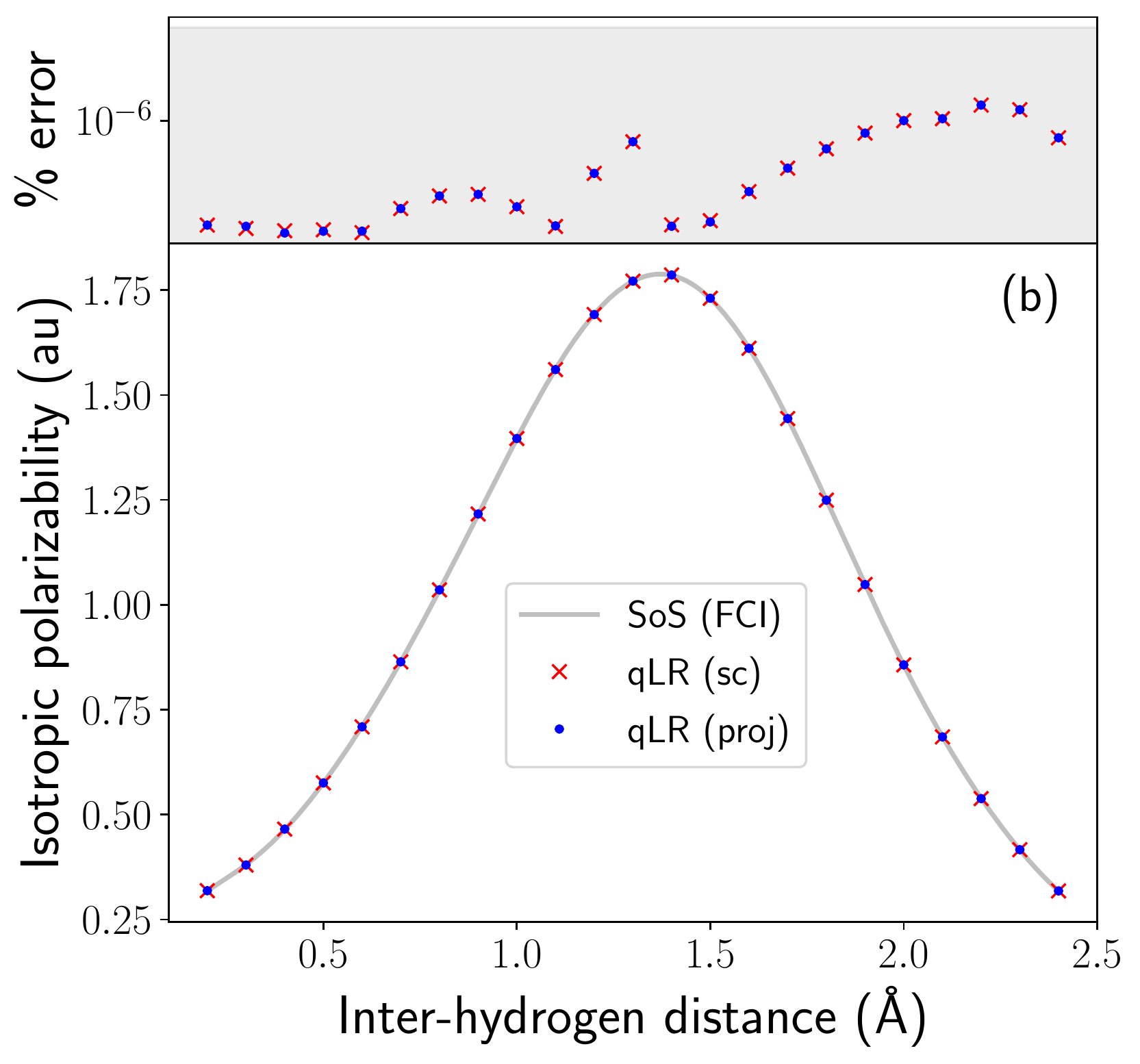}%
        \label{fig:h2_polar}%
        }%
    \caption{(a) Excitation energies ($E_h$) of the H$_2$ molecule calculated using \texttt{q-sc-EOM} and \texttt{q-proj-EOM} approaches as a function of the inter-hydrogen distance and (b) Isotropic dynamic electric-dipole polarizability of the H$_2$ molecule calculated at 589 nm using the \texttt{qLR(sc)} and \texttt{qLR(proj)} approaches as a function of the inter-hydrogen distance. The reference \texttt{SoS(FCI)} values are plotted in grey and the deviations from the reference are shown in the upper panels, where the shaded region indicates errors below 0.1 eV in (a) and below 1\% in (b).}
\end{figure*}

% table showing the problem in transition moment
\begin{table*}
\captionsetup{justification=raggedright}
\caption{Excitation energy ($E_{0k}$ in $E_h$), overlap between the ground and excited states ($\langle 0 | k \rangle$) and the transition dipole moment in the z direction ($\langle 0 | \mu_{z} | k \rangle$, $e\ a_0$) for the excited states of the H$_2$ molecule at the bond length of 0.7 \AA~obtained using \texttt{FCI}, \texttt{q-sc-EOM}, \texttt{q-proj-EOM} and \texttt{qEOM} approaches with the STO-3G basis set.}
\vspace*{5mm}
\begin{ruledtabular}
\begin{tabular}{l|ccc|ccc|ccc|ccc|}
%{} & {} & {} & {} & {} & {} & {} & {} & {} & {} & {} & {} & {} \\ % blank line
\multicolumn{1}{c}{\texttt{States}} &  \multicolumn{3}{c}{\texttt{FCI}} & \multicolumn{3}{c}{\texttt{q-sc-EOM}} & \multicolumn{3}{c}{\texttt{q-proj-EOM}} & \multicolumn{3}{c}{\texttt{qEOM}}\\
\hline
{} & {} & {} & {} & {} & {} & {} & {} & {} & {} & {} & {} & {} \\ % blank line
{} &  \multicolumn{1}{c}{E$_{0k}$}  & \multicolumn{1}{c}{$\langle 0 | k \rangle$} & \multicolumn{1}{c|}{$\langle 0 | \mu_{z} | k \rangle$} &  \multicolumn{1}{c}{E$_{0k}$}  &  \multicolumn{1}{c}{$\langle 0 | k \rangle$} & \multicolumn{1}{c|}{$\langle 0 | \mu_{z} | k \rangle$} &
\multicolumn{1}{c}{E$_{0k}$}  &  \multicolumn{1}{c}{$\langle 0 | k \rangle$} & \multicolumn{1}{c|}{$\langle 0 | \mu_{z} | k \rangle$} &
\multicolumn{1}{c}{E$_{0k}$}  &  \multicolumn{1}{c}{$\langle 0 | k \rangle$} & \multicolumn{1}{c|}{$\langle 0 | \mu_{z} | k \rangle$}\\
%{} & {} & {} & {} & {} & {} & {} & {} & {} & {} & {} & {} & {} \\ % blank line
\hline
{} & {} & {} & {} & {} & {} & {} & {} & {} & {} & {} & {} & {} \\ % blank line
{$T_0$} & {0.6577} & {0} & {0}      & {0.6577} & {0} & {0}        & {0.6577} & {0} & {0}       & {0.6577} & {0} & {0}  \\
\hline
{} & {} & {} & {} & {} & {} & {} & {} & {} & {} & {} & {} & {} \\ % blank line
{$S_1$} & {1.0157} & {0} & {1.1441} & {1.0157} & {0} & {1.1441}   & {1.0157} & {0} & {1.1441}  & {1.0157} & {0} & {1.1441} \\
\hline
{} & {} & {} & {} & {} & {} & {} & {} & {} & {} & {} & {} & {} \\ % blank line
{$S_2$} & {1.7195} & {0} & {0}      & {1.7195} & {0} & {0}        & {1.7195} & {0} & {0}       & {1.7195} & {\textcolor{red}{0.1029}} & {\textcolor{red}{-0.1362}}\\
% \cmidrule(l){3-8}
% {} & {} & {} & {} & {} & {} & {} & {}\\ % blank line
% S$_0$ & -1.17101 & 20.8 & 6.8 & 3.3 & 2.5 & -1.6 & 16.0\vspace{0.3 cm} \\
% \hline
% {} & {} & {} & {} & {}  & {} & {} & {}\\ % blank line
% T$_1$ & 11.30 & 0.08  & 0.46 & 0.50 & 0.24 &  0.25  &  -0.13 \vspace{0.2 cm} \\
% S$_1$ & 13.89 & 1.76  & 2.14 & 1.45 & 0.26 &  -0.15 & -0.11 \vspace{0.2 cm} \\
% T$_2$ & 15.25 & 7.84  & 8.23 & 8.15 & -0.23 & -0.31 & -0.60 \vspace{0.2 cm} \\
% S$_2$ & 17.60 & 10.75 & 11.14 & 10.43 & 1.01 & 0.17 &  0.62 \vspace{0.2 cm} \\
\end{tabular}
\end{ruledtabular}
\label{H2_table}
\end{table*}
The excitation energies (EEs) of the three excited states of the H$_2$ molecule using the STO-3G basis set are plotted in Fig.~\ref{fig:h2_EE} for both \texttt{q-sc-EOM} and \texttt{q-proj-EOM} approaches as a function of the inter-hydrogen distance. The reference full configuration interaction (FCI) EE values are plotted in grey, and the deviations in the EE values from the reference for both \texttt{q-sc-EOM} and \texttt{q-proj-EOM} methods are shown in the upper panel. It can be seen that the errors are less than $10^{-12}$ $E_{h}$ across the entire potential energy curve.
This is not surprising since the excitation manifold of singles and doubles used in this work spans a complete space for the H$_2$ molecule and hence, both the methods are exact for a given basis set.
Table~\ref{H2_table} shows the excitation energy ($E_{0k}$), overlap between the ground and excited states ($\langle 0 | k \rangle$) and the transition dipole moment in the z direction ($\langle 0 | \mu_{z} | k \rangle$) of the H$_2$ molecule at a bond length of 0.7 \AA, obtained using \texttt{FCI}, \texttt{q-sc-EOM}, \texttt{q-proj-EOM} and \texttt{qEOM} approaches. One can see that \texttt{q-sc-EOM} and \texttt{q-proj-EOM} approaches yield identical results to \texttt{FCI} but the \texttt{qEOM} method produces an incorrect value of the dipole transition moment for the $S_2$ excited state. The overlap between the ground state and the $S_2$ excited state is non-zero in the \texttt{qEOM} formalism, leading to the spurious value of the dipole transition moment. Rizzo and co-workers have also talked about the issue of non-orthogonality of ground and excited states in the \texttt{qEOM} approach in their work\cite{rizzo2022one}.
It should be noted that both \texttt{q-sc-EOM} and \texttt{q-proj-EOM} approaches satisfy the killer condition which ensures that the ground and excited states are always orthogonal to each other, leading to a reliable and accurate simulation of the excited state properties.
Figure~\ref{fig:h2_polar} plots the dynamic isotropic electric dipole polarizability of the H$_2$ molecule calculated at 589 nm using the \texttt{qLR(sc)} and \texttt{qLR(proj)} approaches as a function of the inter-hydrogen distance. The reference values obtained using the sum-over-states approach utilizing the FCI wavefunction are denoted as \texttt{SoS(FCI)} and are plotted in gray. The absolute percent errors of both the approaches with respect to the reference \texttt{SoS(FCI)} values are shown in a \texttt{log} plot in the upper panel of the figure where the shaded region indicates errors below 1\%. 
The isotropic polarizability is defined as the negative of one-third of the trace of the electric-dipole polarizability tensor.
It can be easily seen that both \texttt{qLR(sc)} and \texttt{qLR(proj)} approaches produce essentially identical results, with errors always less than $10^{-6}$ \%.
\begin{figure}[htp] 
    \subfloat{%
        \includegraphics[width=1.0\linewidth]{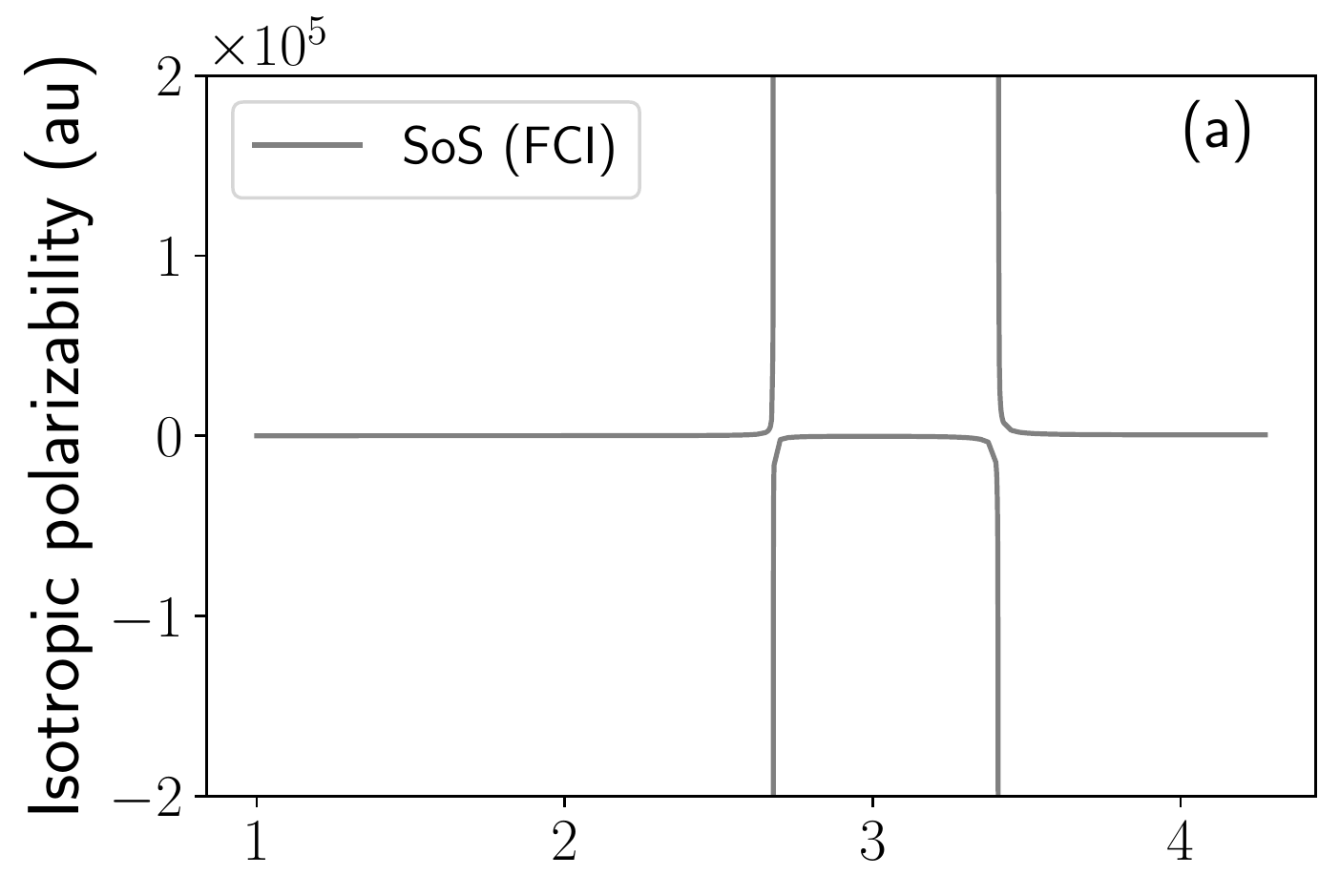}%
        \label{fig:lih_resonance}%
        }%
    \hfill%
    \subfloat{%
        \includegraphics[width=1.0\linewidth]{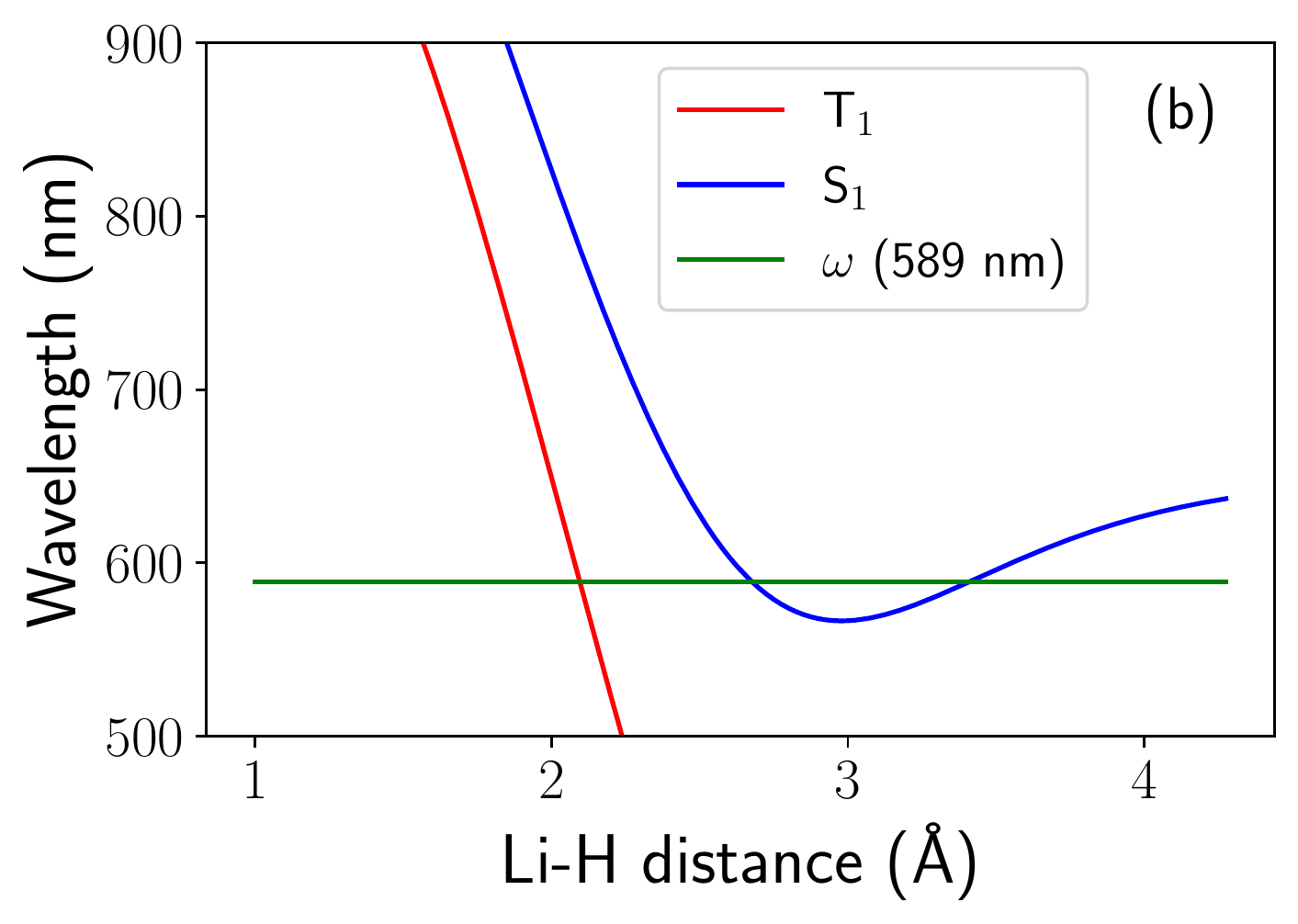}%
        \label{fig:lih_ee_crossing}%
        }%
    \caption{(a) Isotropic electric-dipole polarizability of the LiH molecule  calculated at 589 nm using the \texttt{SoS (FCI)} approach and (b) Excitation energies of the two lowest lying excited states of the LiH molecule, as a function of the Li-H distance.}
\end{figure}

\subsection{LiH}
Figure~\ref{fig:lih_resonance} displays the dynamic electric-dipole polarizability of the LiH molecule calculated at 589 nm using the \texttt{SoS(FCI)} approach as a function of the Li$-$H bond distance. One can see the onset of resonance when the Li$-$H distance is close to 2.7 and 3.4 \AA. Unsurprisingly, the polarizability values approach infinity from positive and negative directions at these two points since the denominator in Eq.~\eqref{SoS} becomes an infinitesimal quantity with both positive and negative signs around the point of resonance. One can verify this from fig.~\ref{fig:lih_ee_crossing} where the excitation energies of the two lowest-lying excited states of the LiH molecule are plotted as a function of the Li$-$H distance. It can be seen that the excitation energy of the first singlet excited state ($S_1$) is equal to the frequency of light (589 nm) around 2.7 and 3.4 \AA. Of course, the triplet excited state is optically forbidden and does not contribute to the polarizability as the ground state is a singlet, resulting in a zero dipole transition moment. One can describe response properties in near-resonance regions through the help of damped response theory~\cite{kristensen2009quasienergy}. However, this is beyond the scope of the current work where we are mostly concerned with calculation of response properties in non-resonant regions. 
\begin{figure*}[tbp]
    \centering
    \captionsetup{justification=raggedright,singlelinecheck=false}
    \subfloat{
        \includegraphics[width = 0.32\textwidth]{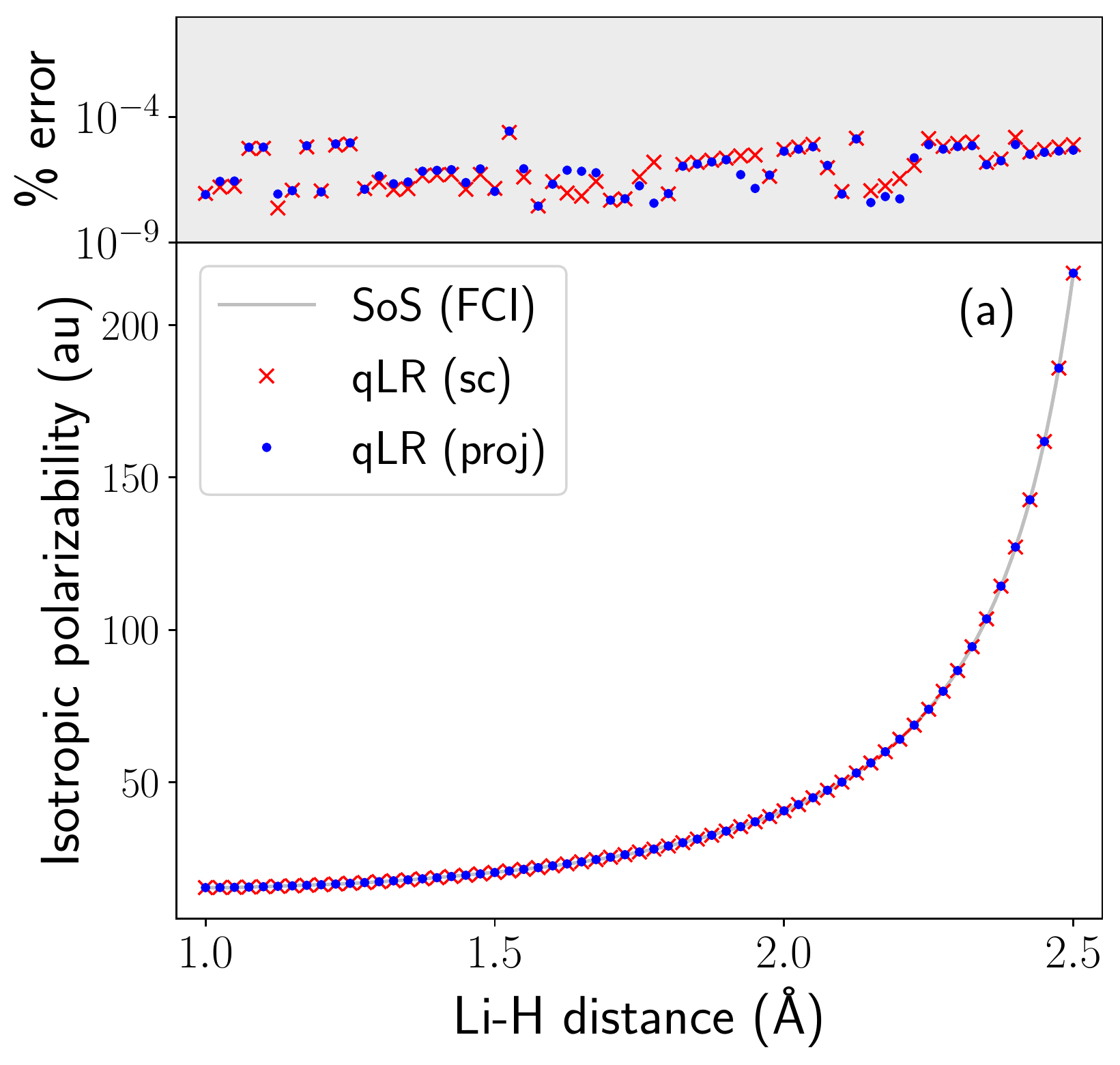}
    }
    \subfloat{
    \includegraphics[width = 0.32 \textwidth]{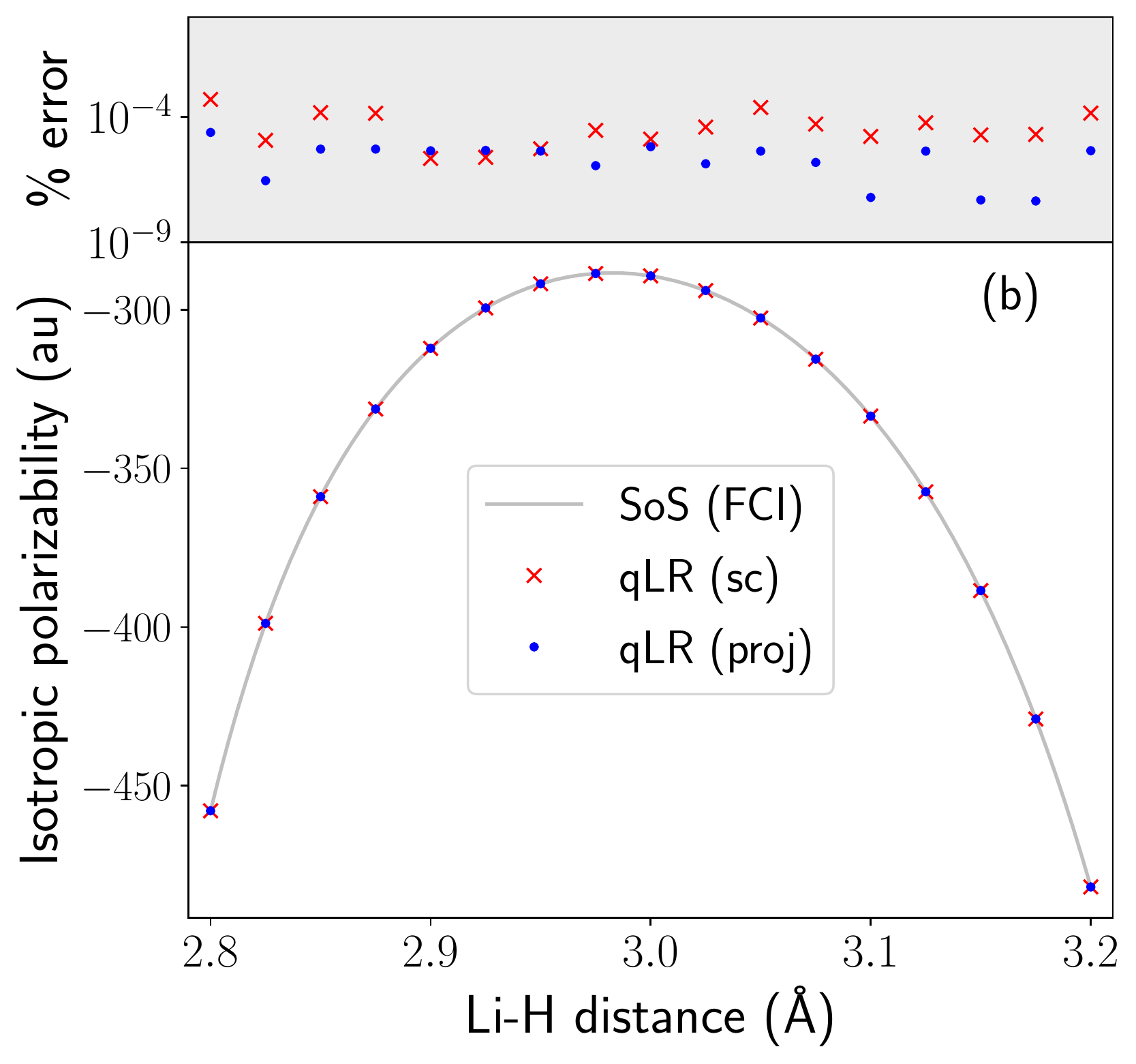}    
    }
     \subfloat{
    \includegraphics[width = 0.32 \textwidth]{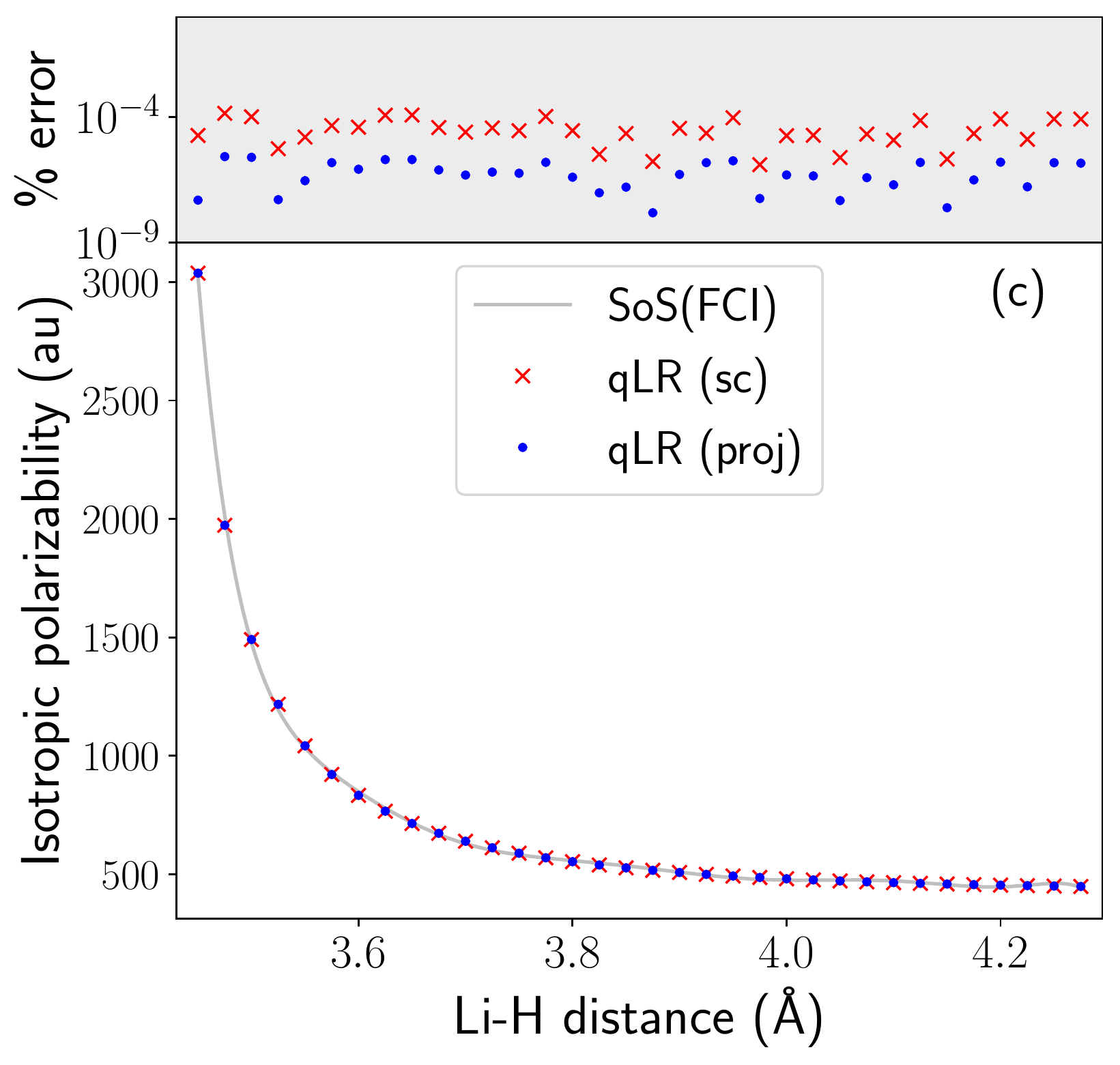}
    }
\caption{Isotropic dynamic electric-dipole polarizability of the LiH molecule calculated at 589 nm using the \texttt{qLR(sc)} and \texttt{qLR(proj)} approaches as a function of the Li$-$H distance in non-resonant regions (a), (b) and (c). The reference \texttt{SoS (FCI)} values are plotted in grey, and the deviations from the reference are shown in the upper panel, where the shaded region indicates errors below 1\%.}
\label{lih_non_resonance}
\end{figure*}
Figure~\ref{lih_non_resonance} plots the dynamic electric-dipole polarizability of the LiH molecule calculated at 589 nm using the \texttt{qLR(sc)} and \texttt{qLR(proj)} approaches as a function of the Li$-$H distance in the three non-resonant regions of the potential energy spectrum.
The reference \texttt{SoS(FCI)} values are plotted in grey and the deviations from the reference are shown in the upper panel, where the shaded region indicates errors below 1\%. It can be seen that the maximal absolute percent error is less than $10^{-3}$\% for both  methods with the errors from the \texttt{qLR(proj)} approach slightly lower in magnitude. 
\subsection{H$_{2}$O}

\begin{figure*}[htp] 
    \centering
     \captionsetup{justification=raggedright,singlelinecheck=false}
    \subfloat{%
        \includegraphics[width=0.47\textwidth]{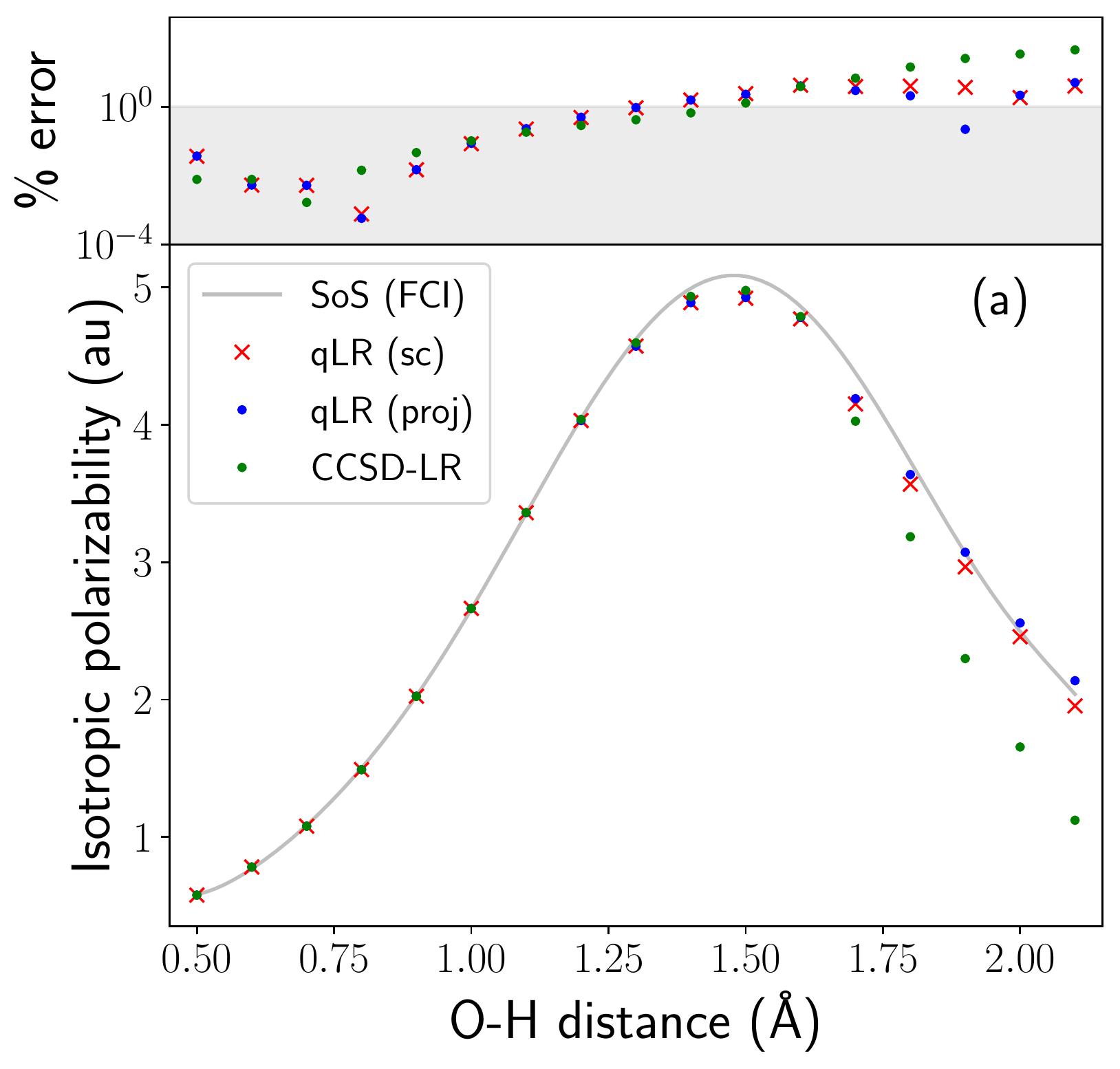}%
        \label{fig:h2o_polar}%
        }%
    \hfill%
    \subfloat{%
        \includegraphics[width=0.47\textwidth]{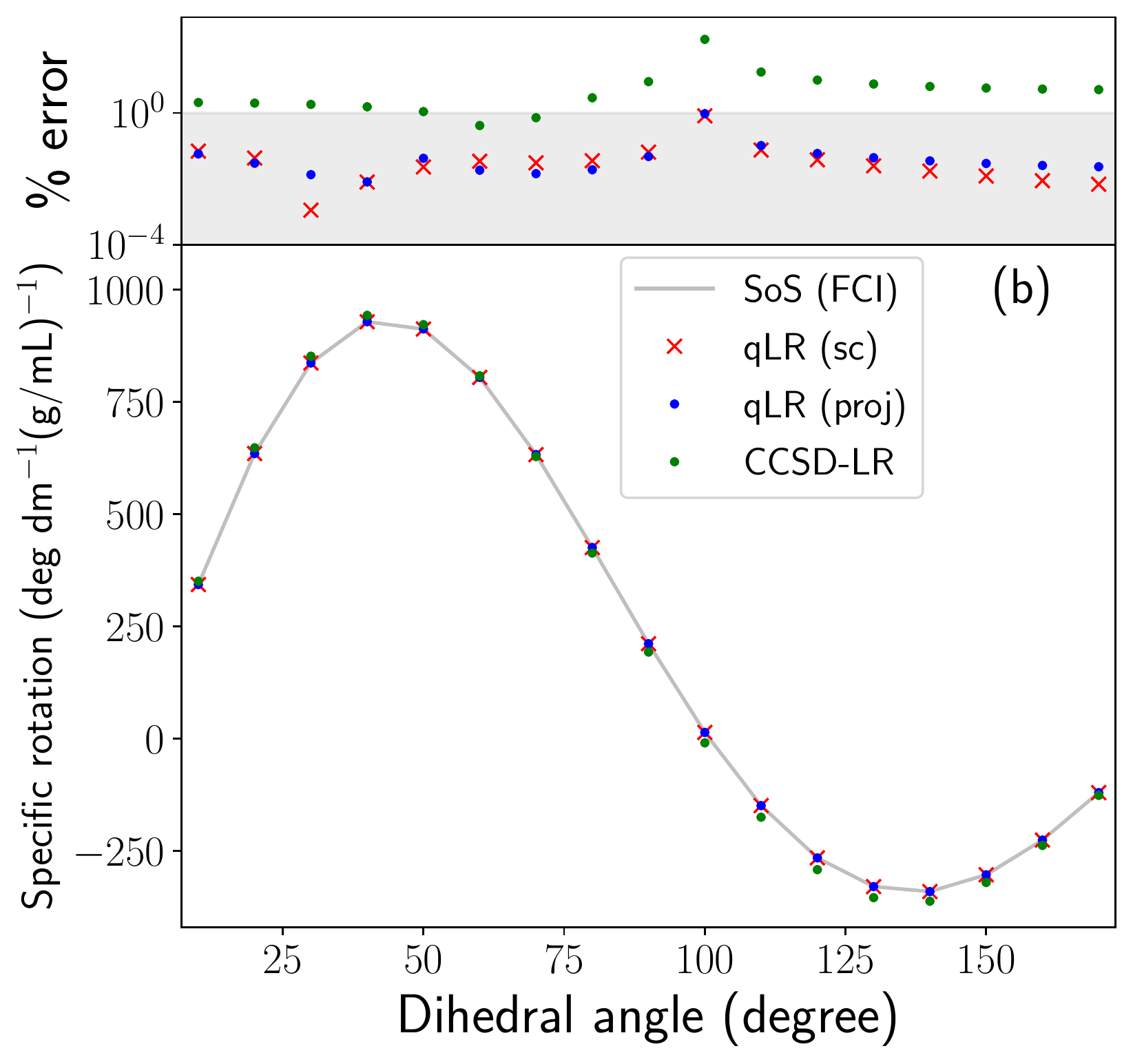}%
        \label{fig:h2_2_spec_rot}%
        }%
    \caption{(a) Isotropic dynamic electric-dipole polarizability of the H$_2$O molecule calculated at 589 nm using the \texttt{qLR(sc)}, \texttt{qLR(proj)} and \texttt{CCSD-LR} approaches as a function of the O$-$H bond distance. The reference \texttt{SoS(FCI)} values are plotted in grey and the deviations from the  reference are shown in the upper panel, where the shaded region indicates errors below 1\%. (b) Specific rotation of the helical ${(\text{H}_{2}})_2$ molecule calculated at 589 nm using the \texttt{qLR(sc)}, \texttt{qLR(proj)} and \texttt{CCSD-LR} approaches as a function of the H$-$H$-$H$-$H dihedral angle. The reference \texttt{SoS(FCI)} values are plotted in grey and the deviations from the reference are shown in the upper panel, where the shaded region indicates errors below 1\%.}
\end{figure*}

Figure~\ref{fig:h2o_polar} plots the dynamic electric-dipole polarizability of the H$_2$O molecule calculated at 589 nm using the quantum (\texttt{qLR(sc)}, \texttt{qLR(proj)}) and classical (\texttt{CCSD-LR}) approaches as a function of the O$-$H bond distance. The reference values (\texttt{SoS(FCI)}) are plotted in grey and the absolute percent errors of all the three approaches with respect to the reference are shown in a \texttt{log} plot in the upper panel of the figure where the shaded region indicates errors below 1\%. One can see that when the O-H bond distance is less than 1.5 \AA, the errors produced by both quantum and classical approaches are close to 1\%. This is due to the fact that this region of the potential energy curve is characterized by weak electron correlation effects. As the O$-$H bond distance increases, strong correlation effects become  dominant and errors start to increase. For example, at an O$-$H bond distance of 2.1 \AA, the errors from \texttt{qLR(sc)}, \texttt{qLR(proj)}, and \texttt{CCSD-LR} methods are close to 4\%, 5\% and 45\%, respectively. Thus, the quantum equation-of-motion approaches result in an order of magnitude reduction in the absolute percent error compared to the classical \texttt{CCSD-LR} method. This is due to the difference in the quality of the underlying ground-state wavefunction. The \texttt{ADAPT-VQE} procedure produces a ground-state wavefunction of similar quality to the \texttt{FCI} wavefunction, while the \texttt{CCSD-LR} method utilizes a \texttt{CCSD} ground-state wavefunction, which provides a very poor description of the electronic structure problem in the strong-correlation domain. It should be noted that all three approaches utilize only singles and doubles excitation operators to describe the time evolution of the ground-state wavefunction under an external time-dependent perturbation. Thus, we can reduce the errors in the quantum-response based approaches by including only a small set of higher-order excitation operators in the parametrization procedure due to the superior quality of the ground-state wavefunction. The left-hand plot in Fig.~\ref{fig:h2o_OS_eq} compares the UV-Vis absorption spectra of the H$_2$O molecule at equilibrium geometry generated using \texttt{FCI}, \texttt{q-sc-EOM} and \texttt{q-proj-EOM} approaches. The spectra produced by \texttt{q-sc-EOM} and \texttt{q-proj-EOM} are indistinguishable from one another and are in qualitative agreement with the \texttt{FCI} results.
\subsection{${(\text{H}_{2}})_2$}
\begin{figure*}[htp] 
    \centering
    \subfloat{%
        \includegraphics[width=0.47\textwidth]{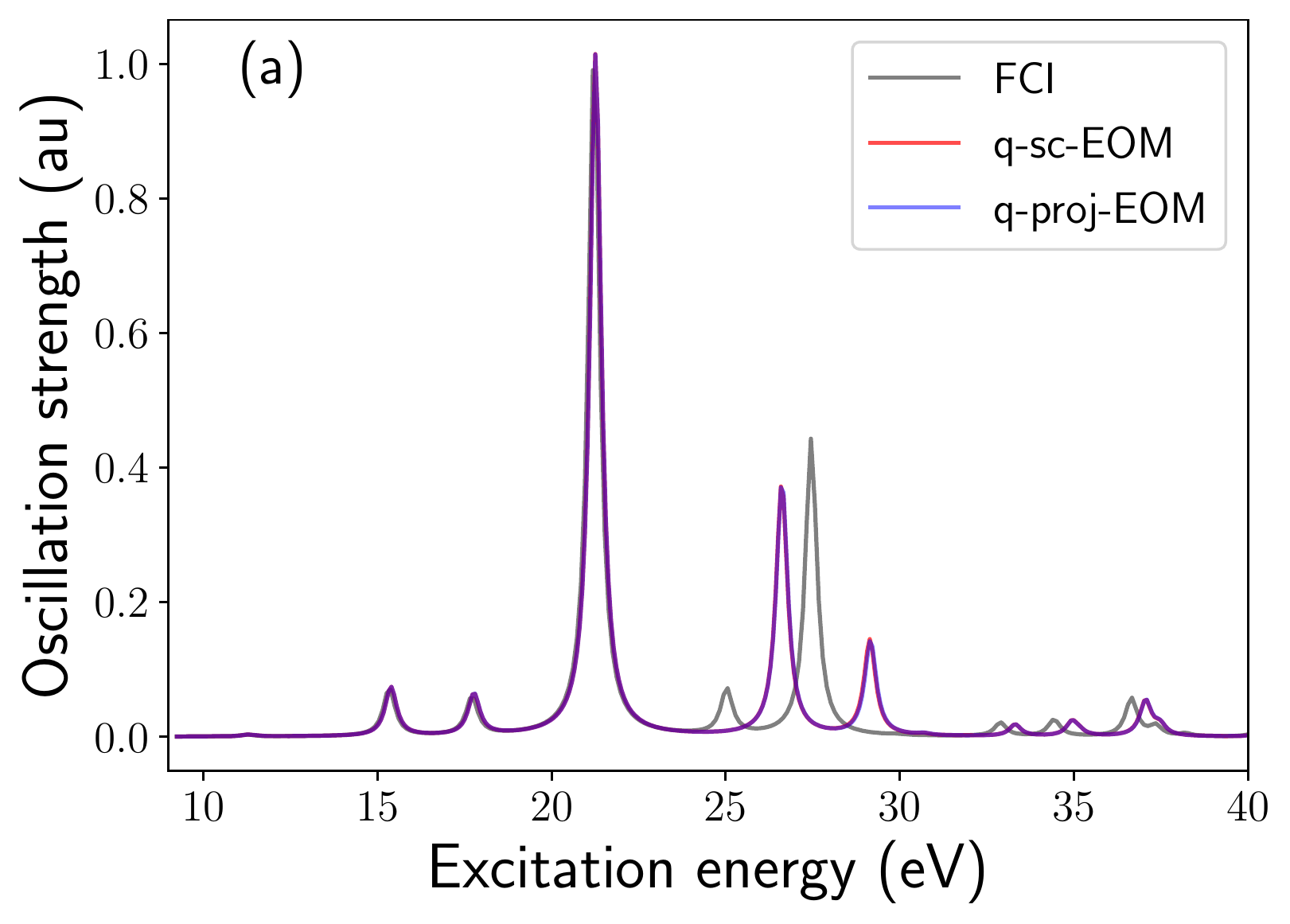}%
        \label{fig:h2o_OS_eq}%
        }%
    \hfill%
    \subfloat{%
        \includegraphics[width=0.47\textwidth]{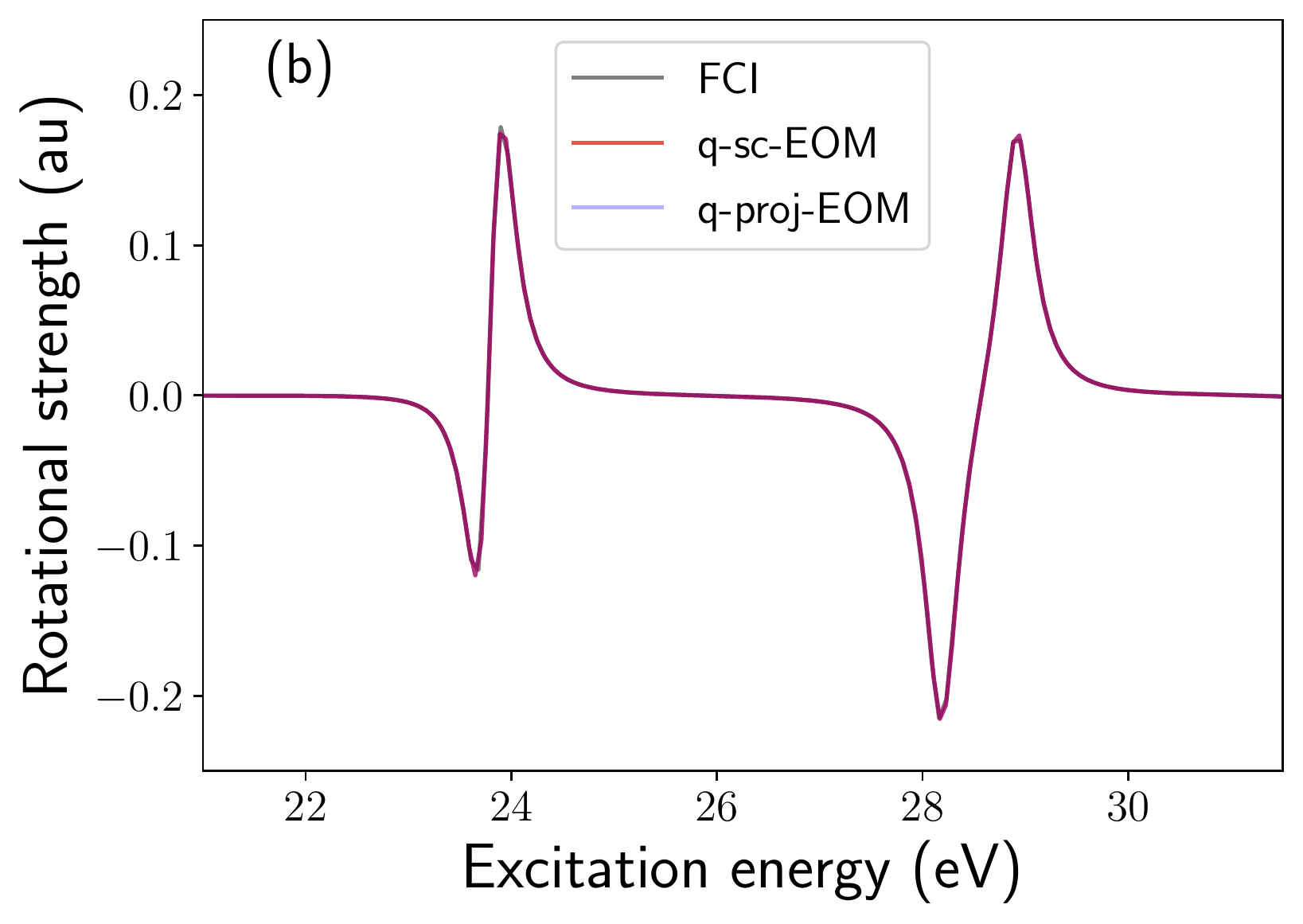}%
        \label{fig:h4_chiral_RS_eq_100}%
        }%
    \caption{(a) UV-Vis spectra of the H$_2$O molecule (equilibrium geometry) and (b) ECD spectra of the ${(\text{H}_{2}})_2$ molecular system (dihedral angle = 100\textdegree), calculated using the \texttt{FCI}, \texttt{q-sc-EOM} and \texttt{q-proj-EOM} approaches.}
\end{figure*}

The magnitude of the optical rotation of a chiral molecule is reflective of its detailed molecular structure and also depends on the frequency of the incident light. Optical rotation, when normalized for path length (dm) and concentration (g/mL), gives the specific rotation [deg dm$^{-1}$(g/ml)$^{-1}$] of the chiral medium. Figure~\ref{fig:h2_2_spec_rot} plots the specific rotation of the ${(\text{H}_{2}})_2$ molecular system calculated at 589 nm using both quantum and classical methods (just like the H$_2$O molecule) as a function of the H$-$H$-$H$-$H dihedral angle. One can see that both the classical (\texttt{CCSD-LR}) and quantum methods produce results which are in qualitative agreement with the reference \texttt{SoS(FCI)} values. However, the errors produced by the classical \texttt{CCSD-LR} method are much larger than those of the quantum approaches across the entire dihedral angle curve. 
For example, when the dihedral angle is equal to 100\textdegree, the errors in \texttt{qLR(sc)}, \texttt{qLR(proj)} and \texttt{CCSD-LR} approaches are 0.8\%, 0.9\% and 170\%, respectively. Furthermore, unlike the quantum approaches, the specific rotation curve produced by the \texttt{CCSD-LR} approach changes sign earlier compared to the reference curve. For example, the values of specific rotation at  100\textdegree \hspace{0.01 in} calculated using \texttt{SoS(FCI)}, \texttt{qLR(sc)}, \texttt{qLR(proj)} and \texttt{CCSD-LR} approaches are 13.8\textdegree, 13.7\textdegree, 13.6\textdegree \hspace{0.01 in} and -9.6\textdegree, respectively. 
It should be noted that the most important criterion for a specific rotation calculation is getting the sign correct. Thus, the quantum approaches offer a clear advantage over their classical counterpart in this regard. 
Absorption spectra (ECD, VCD) of chiral molecules can shed more light on the relationship between molecular structure and the associated optical activity. 
Figure~\ref{fig:h4_chiral_RS_eq_100} compares the ECD absorption spectrum of the ${(\text{H}_{2}})_2$ molecular system (H$-$H$-$H$-$H dihedral angle = 100\textdegree) generated using \texttt{FCI}, \texttt{q-sc-EOM} and \texttt{q-proj-EOM} approaches. Just like the UV-Vis spectra of the H$_2$O molecule, both \texttt{q-sc-EOM} and \texttt{q-proj-EOM} approaches produce identical spectra, which are in perfect agreement with the \texttt{FCI} values.

\subsection{Noise analysis}
\begin{figure}[t]
\includegraphics[width=1.0\linewidth]{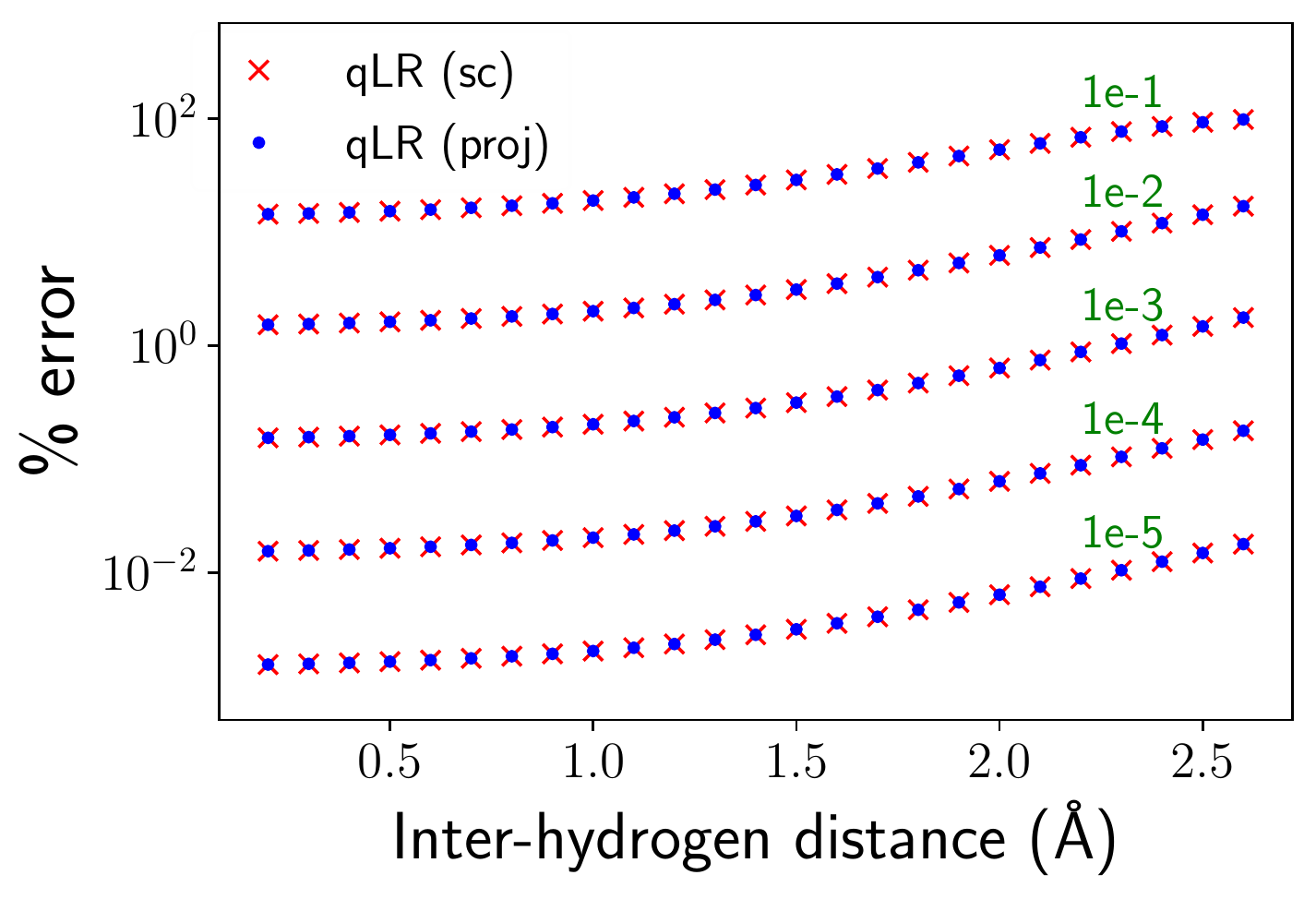}
\captionsetup{justification=raggedright,singlelinecheck=false}
\caption{Percent error in isotropic electric dipole polarizability of the H$_2$ molecule at 589 nm for different values of errors (shown in green) in the optimized ground state parameter, as a function of the inter-hydrogen distance for the \texttt{qLR(sc)} and \texttt{qLR(proj)} approaches.}
\label{h2_polar_gate_error}
\end{figure} 

\begin{figure}[t]
\includegraphics[width=1.0\linewidth]{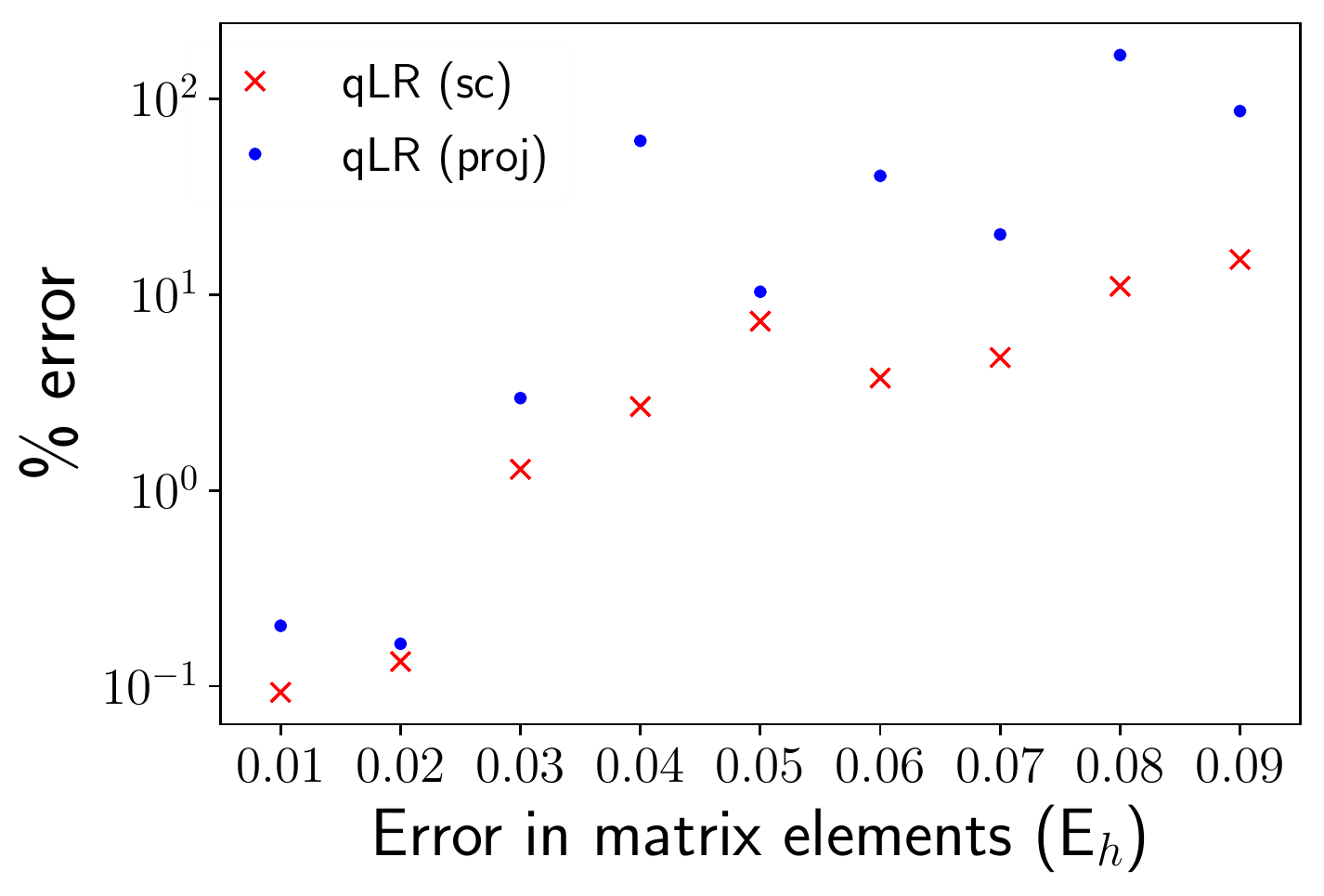}
\caption{Percent error in isotropic electric dipole polarizability of the H$_4$ molecular system at 589  for the \texttt{qLR(sc)} and \texttt{qLR(proj)} approaches.}
\label{h4_polar_shot_error}
\end{figure} 

\begin{figure*}[htp] 
    \centering
    \subfloat{%
        \includegraphics[width=0.47\textwidth]{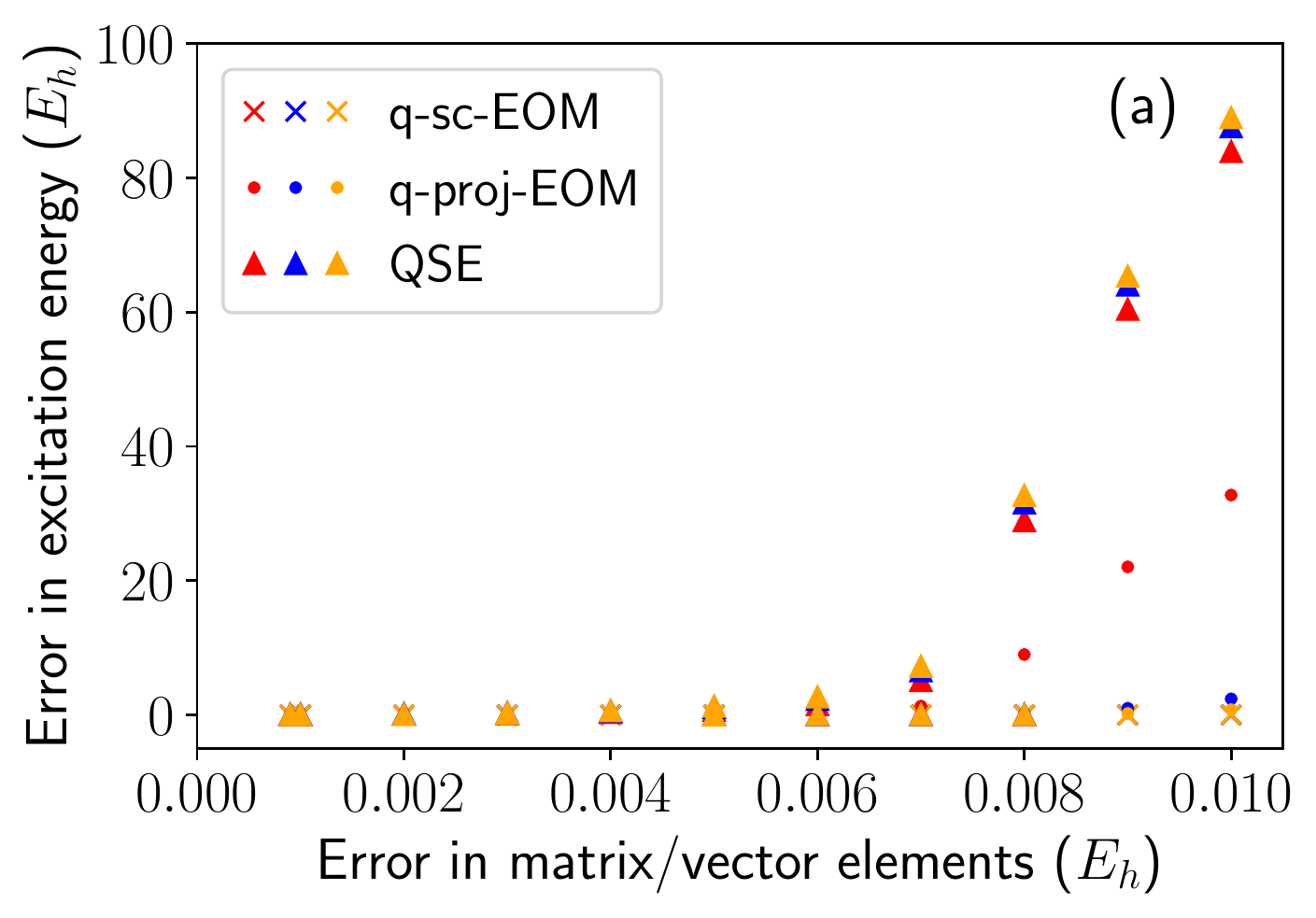}%
        \label{fig:h6_EE_noise}%
        }%
    \hfill%
    \subfloat{%
        \includegraphics[width=0.47\textwidth]{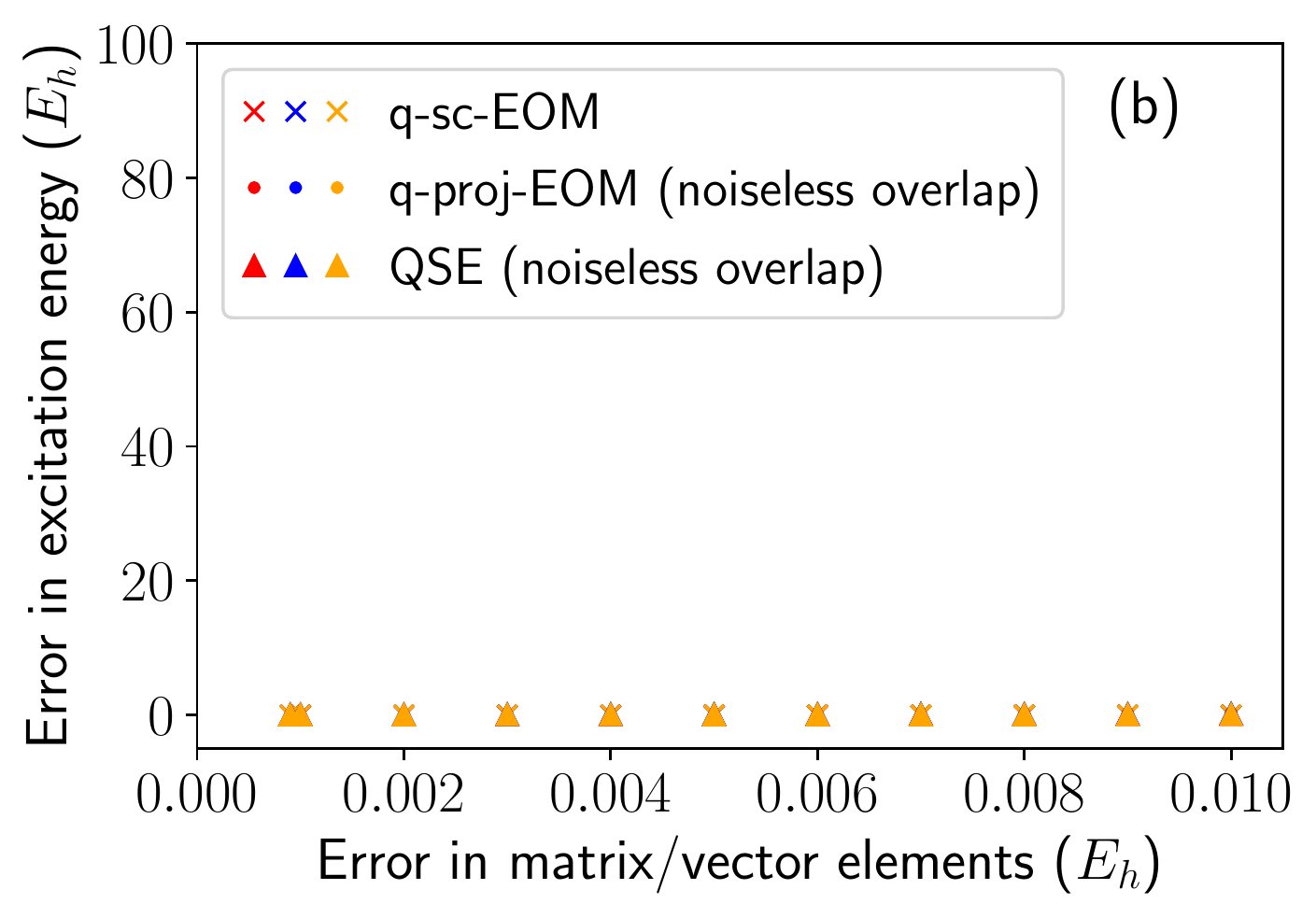}%
        \label{fig:h6_EE_noise_exact_overlap}%
        }%
    \caption{Sensitivity of excitation energies calculated as a function of errors in matrix elements in \texttt{q-sc-EOM}, \texttt{q-proj-EOM} and \texttt{QSE}. In (a),  errors are introduced in all matrices that are expected to be measured on a quantum computer. In (b), errors in the overlap matrix were not introduced in both \texttt{q-proj-EOM} and \texttt{QSE} approaches.}
\end{figure*}

In this subsection we study the stability of qLR formalism to noise.
We first investigate the propagation of errors from the ground state VQE calculation to the isotropic electric-dipole polarizability of the H$_2$ molecule by introducing an error $\epsilon$ to the ground state parameters ($\hat{\sigma}$), followed by a perturbative noise analysis to study the robustness of the above proposed algorithms for statistical errors (can be related to shot noise).
Errors in the ground state amplitudes  ($\hat{\sigma}$) in any physically inspired ansatz can impact the calculated response properties. 
Figure~\ref{h2_polar_gate_error} plots the absolute percent error in the isotropic electric-dipole polarizability of the H$_2$ molecule at 589 nm for different values of the errors (shown in green) in the optimized ground state parameter, as a function of the inter-hydrogen distance for the \texttt{qLR(sc)} and \texttt{qLR(proj)} approaches using the STO-3G basis set.
It can be seen that the percent errors in the isotropic electric-dipole polarizability are higher for every error value ($10^{-1}$ to $10^{-5}$) at larger inter-hydrogen distances. For the induced error of $10^{-3}$ in the ground state amplitudes, the percent error is always less than 1\%; while for $10^{-2}$, the percent error stays always below 10\%. 
In the perturbative noise analysis, we introduce an error from a uniform distribution within an error range (x axis) to each element of matrices $M^{\text{proj}}$, $V^{\text{proj}}$ and vector $Z^{\text{proj}}$ in \texttt{q-proj-EOM} and $M^{\text{sc}}$ and $Z^{\text{sc}}$ in \texttt{q-sc-EOM}~\cite{asthana2022equation}.

Figure~\ref{h4_polar_shot_error} plots the absolute percent error in the isotropic electric-dipole polarizability as a function of the error bounds for H$_4$ in a square planar geometry with a bond length of 1.5 \AA.
 Each data point is an average over 10,000 separate calculations with randomly selected noise within the given bounds shown on the x-axis. One can see that the percent errors in the \texttt{qLR(proj)} approach can be much larger than the ones obtained in the \texttt{qLR(sc)} method. However, it cannot be concluded that this trend will be true for a general molecular system. 
 % { \color {red} One can see that the percent errors in both the approaches rise sharply when the strength of perturbation increases from 0.02 to 0.03, illustrating the high sensitivity of response properties to noise.
 %The max percent error for the \texttt{qLR(sc)} and 
 %\texttt{qLR(proj)} approaches comes out to be around 15\% and 167\%, respectively. Although the percent errors for the \texttt{qLR (proj)} approach is always higher than the \texttt{qLR (sc)} in our analysis, it cannot be concluded that this will be  true for a general molecular system.}
 
 We also carried out a noise sensitivity analysis of  state-specific response properties such as excitation energies computed using the quantum equation-of-motion framework and compared it with the \texttt{QSE} approach.
 In figure~\ref{fig:h6_EE_noise}, we depict the sensitivity of the excitation energies of three lowest-lying excited states of a linear H$_6$ molecular system with a bond distance of 4 \AA, computed using \texttt{q-proj-EOM}, \texttt{q-sc-EOM} and \texttt{QSE} approaches, employing the same perturbative noise formalism as discussed above.
  %Noise is added to each element measured on a quantum computer in form of a perturbative error taken from a uniform distribution within a bound (on the $x$ axis).
  One can see that both \texttt{QSE} and \texttt{q-proj-EOM} methods are more sensitive to the noise compared with the \texttt{q-sc-EOM} formalism. Furthermore, the errors in the \texttt{q-proj-EOM} approach are lower than that of the \texttt{QSE} approach.
  It should be noted that the overlap matrix must be measured on a quantum computer in both \texttt{QSE} and \texttt{q-proj-EOM} approaches. The measured overlap matrix with noise is then inverted to form an eigenvalue equation, a process that is sensitive to noise as discussed in Ref.~\cite{asthana2022equation}.
  In figure~\ref{fig:h6_EE_noise_exact_overlap}, where we artificially eliminate all the noise in the overlap matrices of \texttt{QSE} and \texttt{q-proj-EOM} approaches, one can see very similar noise sensitivity of all three approaches.
  This demonstrates that the noise sensitivity of \texttt{q-proj-EOM} and \texttt{QSE} is a result of measuring the noisy overlap matrix. The overlap matrix in the \texttt{q-sc-EOM} approach on the other hand is exactly known (identity matrix), making this formalism quite noise-resilient. In future work, we plan to employ quantum error mitigation strategies developed for VQE based algorithms\cite{error_mitigation_symmetry_2019,temme2017error,kandala2017hardware,koczor2021exponential}, in conjunction with the \texttt{q-sc-EOM} and \texttt{qLR(sc)} formalisms, to bring down the errors even further.
\section{Conclusions}\label{conclusions}
In this paper, we developed a new protocol for evaluating molecular response properties on near-term quantum computers based on the linear-response framework, named as the quantum linear-response (\texttt{qLR}) theory.
Inspired by the recent work~\cite{asthana2022equation, fan2021equation, liu2022quantum}, we make use of Mukherjee's self-consistent~\cite{prasad1985some} (\texttt{sc}) and Surj\a'an's projected~\cite{szekeres2001killer} (\texttt{proj}) excitation operator manifolds in conjunction with the \texttt{qLR} formalism to make sure that the ``killer condition'' is always satisfied. 
The two proposed formalisms, namely, \texttt{qLR(sc)} and \texttt{qLR(proj)}, 
have been used for the evaluation of dipole polarizabilities and specific rotations of small molecular systems using the ground-state wavefunction obtained through the fermionic \texttt{ADAPT-VQE} algorithm.
We further test the newly developed methods, along with the analogous quantum equation-of-motion (\texttt{qEOM}) variants (\texttt{q-sc-EOM} and \texttt{q-proj-EOM}) to evaluate state-specific response properties like excitation energies, oscillator strengths, and rotational strengths, which were then used to generate UV-Vis and ECD spectra.
Compared to the classical CCSD linear-response (\texttt{CCSD-LR}) theory, we find the quantum methods (without noise) significantly improve the accuracy of response properties near the strong correlation regions due to the better quality of the ground state wavefunction obtained from the quantum approaches.
For example, for the chiral (H$_2$)$_2$ molecular system studied in this work, the specific rotation curve generated by the \texttt{CCSD-LR} theory changes sign earlier than the reference curve. This results in a wrong sign of the specific rotation at some geometric configurations.
In contrast, the \texttt{qLR} approaches get the correct sign of specific rotation at every geometric configuration, with much smaller errors compared to the reference values. Furthermore, in the case of polarizabilities of the H$_2$O molecule, the \texttt{qLR} approaches produced an order of magnitude reduction in the errors compared to the \texttt{CCSD-LR} method in strongly correlated regions of the potential surface.
Since response properties can be quite sensitive to the quality of basis sets, we envision a combination of the \texttt{qLR} approach with the transcorrelated Hamiltonian formalism~\cite{kumar2022quantum} in the future to obtain highly accurate properties using small basis sets.
Through this work, we demonstrated that quantum simulation of response properties using near-term quantum computers can be useful in chemical sciences if the effects of noise are mitigated sufficiently.

\section{Acknowledgement}
A.K., Y.Z., L.C., S.T., and P.A.D.\ thank the support from the Laboratory Directed Research and Development (LDRD) program of Los Alamos National Laboratory (LANL) under project number 20200056DR. LANL is operated by Triad National Security, LLC, for the National Nuclear Security Administration of the U.S. Department of Energy (contract no. 89233218CNA000001).
A.A. and N.J.M.\ would like to acknowledge  U.S. Department of Energy Award No.  DE-SC0019199.  T.D.C.\ was supported by the U.S. National Science Foundation (grant CHE-2154753). 
\appendix
\section*{Appendix}\label{Appendix}
\addcontentsline{toc}{section}{Appendices}
\renewcommand{\thesubsection}{\Alph{subsection}}
%\section{Appendix}\label{Appendix}
%\numberwithin{equation}{section}
\subsection{Expectation value picture of response functions}\label{Exp_value}
The first-order component of the Hamiltonian can be decomposed into the Fourier components as
\begin{equation}\label{Fourier}
    \hat{H}^{(1)}(t)=\int\limits_{-\infty}^{\infty} \mathrm{d} \omega \hat{H}^{(1)}(\omega) e^{-\mathrm{i} \omega t}.\tag{A.1}
\end{equation}
In the spirit of perturbation theory, the wavefunction can be expanded in different orders of the perturbation as
\begin{align}\label{order_expansion}
&\ket{\Psi_{0}(t)} = \ket{\Psi_{0}^{(0)}} + \ket{\Psi_{0}^{(1)}(t)} + \ket{\Psi_{0}^{(2)}(t)} + \ldots ,\tag{A.2}\nonumber
\end{align}
where the time-dependent perturbed wavefunction of a given order can be represented in terms of the Fourier transforms of their frequency-dependent counterparts, just like ~\eqref{Fourier},
\begin{align}
&\ket{\Psi_{0}^{(1)}(t)} = \int\limits_{-\infty}^{\infty}\mathrm{d} \omega e^{-\mathrm{i} \omega t} \ket{\Psi_{0}^{(1)}(\omega)} \tag{A.3}\\ 
&\ket{\Psi_{0}^{(2)}(t)} = \int_{-\infty}^{\infty} d \omega_{1} \int_{-\infty}^{\infty} d \omega_{2}
\hspace{0.02 in}e^{-i( \omega_{1}+ \omega_{2})t}
\ket{\Psi_{0}^{(2)}(\omega_1,\omega_2)}.\nonumber
\end{align}

% Here, the $n^{\text{th}}$ order wavefunction in frequency domain, $\ket{\Psi_{0}^{(n)}(\omega_1,\omega_2,..,\omega_n)}$, models the interaction of the molecular system with ${n}$ photons at their corresponding frequencies.

One can arrive at the closed expressions for 
response functions of different orders by expanding the time-dependent expectation value of a time-independent Hermitian operator $\hat{X}$ in orders of the perturbation $\hat{H}^{(1)}(t)$, e.g.,
%\begin{align}
\begin{equation}
\resizebox{0.999\hsize}{!}{$ \begin{split}
    &\langle\hat{X}\rangle(t)=
    \langle\hat{X}\rangle^{(0)}+\int_{-\infty}^{\infty} d \omega\left\langle\left\langle\hat{X} ; \hat{H}^{(1)}(\omega)\right\rangle\right\rangle e^{-i \omega t} \hspace{0.02 in} + \\ 
    &\frac{1}{2} \int_{-\infty}^{\infty} d \omega_{1} \int_{-\infty}^{\infty} d \omega_{2}\left\langle\left\langle\hat{X} ; \hat{\text{H}}^{(1)}\left(\omega_{1}\right), \hat{\text{H}}^{(1)}\left(\omega_{2}\right)\right\rangle\right\rangle e^{-i( \omega_{1}+ \omega_{2})t} \\&+\ldots\hspace{0.02 in},\nonumber
    \end{split}$}
    \tag{A.4}
\end{equation}   
where $\langle\hat{X}\rangle^{(0)}$ is the expectation value of the $\hat{X}$ operator with respect to the unperturbed time-independent ground-state wavefunction, $\left\langle\left\langle\hat{X} ; \hat{H}^{(1)}(\omega)\right\rangle\right\rangle$
and $\left\langle\left\langle\hat{X} ; \hat{\text{H}}^{(1)}\left(\omega_{1}\right), \hat{\text{H}}^{(1)}\left(\omega_{2}\right)\right\rangle\right\rangle$ refer to the linear and quadratic response functions, respectively, and so on. 
A response function of a given order in the presence of a given external field is associated with a specific physical phenomenon. 
For example, if $\hat{X}$ is the electric dipole operator, $\vec{\mu}$, and $\hat{H}^{(1)}(\omega)$ corresponds to an oscillating electric field, then the associated linear response function refers to the negative of the dynamic dipole polarizability ($\alpha$) of the molecule, e.g.,
\begin{equation}
\left\langle\left\langle\hat{X} ; \hat{H}^{(1)}(\omega)\right\rangle\right\rangle=\langle\langle\vec{\mu} ; \vec{\mu}\rangle\rangle(\omega)=-\alpha(\omega).\tag{A.5}
\end{equation}
%Similarly, higher-order response functions correspond to first- and higher-order hyperpolarizabilities, where the various frequencies, $\omega_i$ may be chosen to represent specific combinations or overtones. 
If the perturbation is a magnetic field instead, then the imaginary part of the associated linear response function gives the Rosenfeld tensor, the trace of which is related to the optical rotation of the molecular system,
\begin{equation}
\left\langle\left\langle\hat{X} ; \hat{H}^{(1)}(\omega)\right\rangle\right\rangle=\langle\langle\vec{\mu} ; \vec{m}\rangle\rangle(\omega)= G^{\prime}(\omega),\tag{A.6}
\end{equation}
where $\vec{m}$ corresponds to the magnetic moment operator.\\
\subsection{Equation of motion (EOM) theory}\label{EOM}
The EOM formalism involves explicit evaluation of the excited states and the corresponding excitation energies and makes use of the sum of states approach (equation~\ref{SoS}) to calculate response properties.
The wavefunction for the $\text{k}^{\text{th}}$ excited state ($\ket{\Psi_\text{k}}$) can be obtained by the action of a state-transfer operator ($\hat{\mathbb{O}}_\text{k}$) on the ground-state wavefunction ($\ket{\Psi_{\text{0}}}$), 
\begin{align}
   \ket{\Psi_\text{k}} = \hat{\mathbb{O}}_\text{k}\ket{\Psi_{\text{0}}},\tag{B.1}
\end{align}
where $\hat{\mathbb{O}}_\text{k}$ has the same basic form as the time-dependent cluster operator (Eq.~\eqref{R_resp}) in the linear response formalism,
\begin{align}\label{OK}
    % \hat{\mathbb{O}}_{\text{k}} & = \hat{\text{R}}^\text{k}_1 + \hat{\text{R}}^\text{k}_2 + \hat{\text{R}}^\text{k}_3 + \dots
    \hat{\mathbb{O}}_{\text{k}} & = \sum_{i=1}^{N}\hat{R}^\text{k}_i ,\nonumber\\
     \hat{R}^{\text{k}}_i & =\sum_{\mu}\big[A^{\text{k}}_{\mu_{i}} \hat{G}_{\mu_{i}}+B^{\text{k}}_{\mu_i^{\dagger}}\hat{G}_{\mu_i^\dagger}\big],\tag{B.2}
\end{align}
except that the coefficients $A^{\text{k}}_{\mu_{i}}$ and $B^{\text{k}}_{\mu_i^{\dagger}}$ are now time-independent and state-specific.
Assuming the ground-state wavefunction and the state-transfer operator to be exact, the excitation energy associated with the transition from ground to k$^{\text{th}}$ excited state can be obtained through the application of a commutator of the Hamiltonian and the corresponding state-transfer operator on the ground-state wavefunction, which can be written as
\begin{align}
\begin{split}
    [\hat{\text{H}},\hat{\mathbb{O}}_\text{k}]|\Psi_{\text{0}}\rangle=&\hspace{0.05 in}\hat{\text{H}}\hat{\mathbb{O}}_\text{k}|\Psi_{\text{0}}\rangle-\hat{\mathbb{O}}_\text{k}\hat{\text{H}}|\Psi_{\text{0}}\rangle,\\
    =&\hspace{0.05 in}(\text{E}_{\text{k}}-\text{E}_{\text{0}})\hspace{0.02 in}\hat{\mathbb{O}}_\text{k}|\Psi_{\text{0}}\rangle.\label{eommaineq}
    %=&\hspace{0.05 in}E_{0k}\hat{\mathbb{O}}_k|\Psi_{\text{0}}\rangle.\label{eommaineq}
\end{split}
\tag{B.3}
\end{align}
where E$_{\text{0}}$ and E$_{\text{k}}$ are the energies of the ground and the $\text{k}^{\text{th}}$ excited state, respectively. Here, $\hat{\text{H}}$ refers to the molecular Hamiltonian operator in the second-quantized form. 
By projecting the above equation onto the $\text{k}^{\text{th}}$ excited state wavefunction,
and assuming the vacuum annihilation condition holds true (see sec.~\ref{VAC} for details),
one can compute the excitation energy directly as
\begin{align}
\begin{split}
    \text{E}_{\text{0k}}=&\frac{\langle \Psi_{\text{0}}|[(\hat{\mathbb{O}}^\dagger_\text{k}),[\hat{\text{H}},(\hat{\mathbb{O}}_\text{k})]]|\Psi_{\text{0}}\rangle}{\langle \Psi_{\text{0}}|[(\hat{\mathbb{O}}^\dagger_\text{k}),(\hat{\mathbb{O}}_\text{k})]|\Psi_{\text{0}}\rangle}.
\end{split}\tag{B.4}\label{qeomeq}
\end{align}
It can be seen that Eq.~\eqref{qeomeq} has a parametric dependence on the excitation ($A^\text{k}_{\mu_i}$) and de-excitation amplitudes ($B^\text{k}_{\mu_i^\dagger}$). By doing a variational minimization of the equation ($\delta \text{E}_{\text{0k}}$ = 0), one can arrive at the following secular equation to solve for these amplitudes,\\
\begin{align}\label{secular}
\left(\begin{array}{cc}
\mathbf{M} & \mathbf{Q} \\
\mathbf{Q}^{*} & \mathbf{M}^{*}
\end{array}\right)\left(\begin{array}{l}
\mathbf{A}^{\text{k}} \\
\mathbf{B}^{\text{k}}
\end{array}\right)=\text{E}_{\text{0k}}\left(\begin{array}{cc}
\mathbf{V} & \mathbf{W} \\
-\mathbf{W}^{*} & -\mathbf{V}^{*}
\end{array}\right)\left(\begin{array}{l}
\mathbf{A}^{\text{k}} \\
\mathbf{B}^{\text{k}}
\end{array}\right),
\tag{B.5}
\end{align}
where the expression for the matrix elements of $\textbf{M}$, $\textbf{Q}$, $\textbf{V}$ and $\textbf{W}$ are the same as of Eq.~\eqref{MVQW}.
In this work, for computational convenience, we restrict the max rank of excitation and de-excitation operators to two, i.e. \{$\text{i},\text{j}\} \in \{1,2\}$. 
One is able to achieve 
``quantum advantage'' through quantum measurements of these matrix elements since their classical evaluations will have a factorial scaling with respect to the system size for an exact ground-state wavefunction. 

In this formalism, the ground-state wavefunction can be obtained in principle from any popular quantum algorithms. However, we employ the ADAPT-VQE procedure to obtain the ground state wavefunction as it produces compact quantum circuits.
Ollitrault and co-workers~\cite{qEOM2020} developed and implemented this formalism on a quantum computer and named it as the ``qEOM'' method. Once the measurements are done, the generalized eigenvalue equation can be solved to obtain excitation energies, which possess the favorable property of size-intensivity.
Furthermore, from the eigenvectors of Eq.~\eqref{secular}, one can calculate transition moments which can be used to calculate molecular absorption spectra. For example, to generate the ultraviolet-visible (UV-Vis) spectra for a molecular system, one needs both excitation energies and oscillator strengths (OS) corresponding to different excited states where the OS gives the probability of an electric dipole transition from the ground-state to a given excited state. Similarly, one can generate an ECD spectra for chiral molecules by calculating rotational strengths (RS) for different excited states. For the $\text{k}^{\text{th}}$ excited state, these quantities are defined as
\begin{align}
    \textbf{OS}_\text{k} & = \sum_i \frac{2}{3} \text{E}_{\text{0k}} \big[\langle \Psi_{\text{0}} |\vec{\boldsymbol{\mu}}_i| \Psi_{\text{k}}\rangle\langle \Psi_{\text{k}}|\vec{\boldsymbol{\mu}}_i| \Psi_{\text{0}} \rangle\nonumber\\
    \textbf{RS}_{\text{k}} & = \sum_i \big[\langle \Psi_{\text{0}} |\vec{\boldsymbol{\mu}}_i| \Psi_{\text{k}}\rangle\langle \Psi_{\text{k}}|\vec{\mathbf{m}}_i| \Psi_{\text{0}} \rangle
    \tag{B.6}
\end{align}
where $i \in \{x,y,z\}$. Equation~\eqref{secular} has a structure known as generalized Random-Phase approximation. In the context of quantum chemistry, this eigenvalue problem is frequently solved for TDHF or TDDFT methods.  Of course, in TDHF (or TDDFT) approaches,  only rank one excitations and de-excitations are considered and the Hartee-Fock (or Kohn-Sham) determinant is taken as the reference wavefunction. Just like in the case of the TDHF/TDDFT formalism, one can encounter potential numerical issues while solving Eq.~\eqref{secular} as it is not a generalized Hermitian eigenvalue problem~\cite{HIRATA1999291}. One can always reformulate this equation into a generalized Hermitian eigenvalue equation to solve for 
$\textbf{A}_k \pm \textbf{B}_k$ instead, but the $\textbf{M} - \textbf{Q}$ matrix appearing in such a formulation would need to be strictly positive-definite, which might not be always guaranteed. Moreover, one needs to calculate inverse of $\textbf{M} \pm \textbf{Q}$ matrices, which could also cause potential numerical instabilities. One possible way to avoid this problem is by employing a Tam-Dancoff (TDA)~\cite{HIRATA1999291} like approximation and neglecting the de-excitation operators altogether. 
The QSE method does employ the TDA approximation but also includes identity in its operator pool, due to which the excitation energies obtained from QSE are not size-intensive.\\

\bibliography{main}% Produces the bibliography via BibTeX.
\end{document}